\documentclass[aps,prd,preprint,groupedaddress,showkeys]{revtex4-2}

\usepackage{amsmath}
\usepackage{graphicx}
\usepackage{bm}
\usepackage{dcolumn}
\usepackage{latexsym}
\DeclareMathOperator\arctanh{arctanh}

\begin{document}

\title{Electromagnetic wave propagation in general Kasner-like metrics}
\author{Brett Bochner}
\email[]{Brett.D.Bochner@hofstra.edu, brett\_bochner@alum.mit.edu}
\affiliation{Department of Physics and Astronomy, Hofstra University, Hempstead, NY 11549}

\date{\today}

\begin{abstract}
The curved spacetime Maxwell equations are applied to the anisotropically 
expanding Kasner metrics. Using the application of vector identities 
we derive 2\textsuperscript{nd}-order differential wave equations for the 
electromagnetic field components; through this explicit derivation, we find 
that the 2\textsuperscript{nd}-order wave equations are not uncoupled for the 
various components (as previously assumed), but that gravitationally-induced 
coupling between the electric and magnetic field components is 
generated directly by the anisotropy of the expansion. The lack of 
such coupling terms in the wave equations from several prior studies 
may indicate a generally incomplete understanding of the evolution of 
electromagnetic energy in anisotropic cosmologies. Uncoupling the field 
components requires the derivation of a 4\textsuperscript{th}-order wave 
equation, which we obtain for Kasner-like metrics with generalized 
expansion/contraction rate indices. For the axisymmetric Kasner case, 
$(p_{1}, p_{2}, p_{3}) = (1,0,0)$, we obtain exact field solutions 
(for general propagation wavevectors), half of which appear not to have 
been found before in previous studies. For the other axisymmetric Kasner case, 
$\{p_{1}, p_{2}, p_{3}\} = \{(-1/3),(2/3),(2/3)\}$, we use numerical methods 
to demonstrate the explicit violation of the geometric optics 
approximation at early times, showing the physical phase velocity 
of the wave to be inhibited towards the initial singularity, 
with $v \rightarrow 0$ as $t \rightarrow 0$.
\end{abstract}

\keywords{Geometric optics violation; Kasner metric; Cosmology; Curved spacetime Maxwell equations}

\maketitle

\section{Introduction}
\label{SecIntro}
In this paper we study the propagation of electromagnetic waves 
in the homogeneous but anisotropic Kasner metrics. These metrics 
are interesting physically, as exact solutions to the 
Einstein field equations possessing a singularity (for most Kasner metric 
parameters) at $t=0$; and they are interesting cosmologically, as 
building blocks which can be generalized to help construct the 
Mixmaster universe model \citep{BKL}.

With a convenient alignment of our spatial axes, and noting that 
we set the speed of light to unity ($c \equiv 1$) throughout this paper, 
the Kasner metric is written as \citep{MTW}:
\begin{equation}
ds^{2} = -dt^{2} + t^{2 p_{x}} dx^{2} + t^{2 p_{y}} dy^{2} + t^{2 p_{z}} dz^{2} ~, 
\label{KasMetric}
\end{equation}
where the indices are typically assumed to satisfy the conditions: 
\begin{equation}
(p_{x} + p_{y} + p_{z}) = (p_{x}^{2} + p_{y}^{2} + p_{z}^{2}) = 1 ~. 
\label{KasParams}
\end{equation}
These Kasner conditions, if applied, ensure the vacuum nature 
(stress-energy tensor $T^{\mu \nu} = 0$) of the metric.

Modeling the propagation of light as null rays here, using the 
geometric optics approximation \citep{MTW}, would be comparatively 
straightforward; but to study much of the interesting physics, 
a wave treatment using Maxwell's equations is needed. 
For example, it is known that plane wave solutions will in general 
not follow null geodesics in anisotropic spacetimes, except in the 
high frequency (geometric optics) limit; hence effects such as 
birefringence (and other important phenomena) will be missed 
in a null ray treatment \citep{AsenjoHoj}.

Investigations of Maxwell's equations in the Kasner metric, however, have 
proven to generate wave equations that only in rare or special cases have  
analytical solutions. Specifically, the problem of finding exact solutions 
for general wavevectors (all propagation directions), which are good for 
all values of time $t$, only appears achievable (among those cases obeying 
Eq.~\ref{KasParams}) for the $(p_{x}, p_{y}, p_{z}) = (1,0,0)$ case. 
For that case specifically, analytical wave solutions were derived 
in Sagnotti and Zwiebach \cite{SagZwiebachWaves} as linear combinations 
of Bessel functions (i.e., Hankel functions) of purely imaginary order; 
and derivations by other authors \citep[e.g.,][]{OlPetKasner} have 
led to equivalent results. 

For other sets of Kasner indices -- even the other axisymmetric 
vacuum case, $\{p_{x}, p_{y}, p_{z}\} = \{(-1/3),(2/3),(2/3)\}$ -- 
analytical solutions appear to be unavailable, without adopting 
various simplifications. Using a superpotential formalism developed 
by Kegeles and Cohen \cite{KegCohen} for calculating electromagnetic (as well as 
neutrino and gravitational) metric perturbations, this Kasner case 
was studied in a series of papers by different authors 
\citep{DhurVishVeshCohen, PonsMarciSeries, AnsaryFieldSeries}, 
in which analytical solutions were obtained only within early-time 
$(t << 1)$ and late-time $(t >> 1)$ approximations. Alternatively, 
some authors \citep{GoorjianKas,SagZwiebachWaves,OlPetKasner} 
have found solutions to this case (or for more general choices of 
Kasner parameters) by restricting the wave propagation direction 
to be along a single coordinate axis (i.e., along one of the 
principal expansion axes); and some \citep[e.g.,][]{MatWollKasEM} 
have employed combinations of both approximations. 

In some recent papers \citep{EspoScalarKas,AlhoBBwvEqn,OlPetKasner}, 
wave propagation has been studied in the Kasner metric using 
the {\it scalar} wave equation:
\begin{equation}
\Box_{g} \equiv g^{\mu \nu} \nabla_{\mu} \nabla_{\nu} \phi = 0 ~, 
\label{ScalarWaves}
\end{equation}
with the d'Alembertian transformed appropriately for the relevant metric 
being studied. In a lengthier exposition by Petersen \cite{OlPetThesis}, 
it is made clear (for example in their expression, 
``Scalar wave equation for light") that this equation is actually 
being used to study the propagation of {\it electromagnetic} fields 
in the Kasner metric. While this may or may not cause discrepancies 
in some particular study (e.g., deriving 
cosmological redshifts \citep{OlPetKasner,OlPetThesis}), 
it is not a rigorous procedure in general to apply a scalar wave 
equation individually to the various components of a vector field, 
implicitly assuming that they are mutually independent. 

On the contrary, one of the principal findings of our paper -- to be shown via 
explicit step-by-step derivations using Maxwell's equations in curved spacetime -- 
is that one cannot specify {\it homogeneous} 2\textsuperscript{nd}-order 
differential wave equations for all of the electromagnetic fields in the 
anisotropic Kasner metric. (Note that for brevity, we will often use the 
terms ``field components" and ``fields" interchangeably in this paper.) 
Assuming a general wave propagation direction, nonhomogeneous terms appear on 
the right-hand side (RHS) of the equations analogous to Eq.~\ref{ScalarWaves} 
for some of the fields -- even in this charge-free spacetime -- generated 
directly by the anisotropy of the axis expansion rates. The RHS term(s) for one 
given field component are proportional to other electromagnetic field components, 
thus linking the fields together, making them mutually dependent (an effect 
which would be absent in any analysis based upon Eq.~\ref{ScalarWaves}). 
For any two axes expanding/contracting according to different 
Kasner parameters $(p_{i} \ne p_{j})$, and for a wavevector possessing 
nonzero components along both of those axes, the electric $(E-)$ or 
magnetic $(B-)$field in the direction normal to 
that ``$ij$-plane" will experience a driving term from its complementary 
fields within the plane. (For example, if $p_{y} \ne p_{z}$, and wavevector 
components $(k_{y}, k_{z}) \ne 0$, then $E_{x}$ would be driven by nonzero 
$B_{y}$ and/or $B_{z}$; $B_{x}$ would be driven by nonzero $E_{y}$ 
and/or $E_{z}$; and so on.) 

The main effect of these driving terms, is that unlike the situation 
for flat spacetime, the individual electromagnetic field components 
(for general wavevector directions) are {\it not} uncoupled in the 
2\textsuperscript{nd}-order wave equations; instead, one must go on 
to derive 4\textsuperscript{th}-order differential equations in order to 
obtain truly independent wave equations for the six electromagnetic fields. 
The fact that these fields must satisfy 4\textsuperscript{th}-order 
equations means that there are actually {\it four} solutions for each of them, 
in general, instead of two; and previous results in the literature for 
such fields that have specified only two field solutions are 
actually missing half of the solution functions. 

Below, we will derive the 4\textsuperscript{th}-order wave equation (which is 
identical for all six of the electromagnetic fields), obtaining a general expression 
that is valid for any values of the Kasner parameters $(p_{x}, p_{y}, p_{z})$, 
including those ``Kasner-like" metrics not restricted to obey the 
vacuum Kasner conditions (Eq.~\ref{KasParams}). Furthermore, 
re-examining the $(p_{x}, p_{y}, p_{z}) = (1,0,0)$ case, we will find 
all four solutions (including the two already known), group them into the 
two natural polarization states for that axisymmetric metric, 
and obtain the general solution (including all field amplitudes) that completely 
solves the Maxwell equations in the case of a general propagation direction, 
with nonzero wavevector components along all distinctly evolving axes. 

These additional field solutions -- and the field-coupling driving terms 
on the RHS of the 2\textsuperscript{nd}-order wave equations -- have 
not been seen by us to have been derived in any previous work. 
Why these additional terms and field solutions have been missed previously 
(if indeed they have been), is not immediately clear. One likely 
contributing factor, is that given the difficulty of finding solutions for 
general propagation directions, the simplification of restricting the 
wave propagation to lie along one of the principal axes is often made 
almost immediately \citep[e.g.,][]{GoorjianKas,MatWollKasEM,AsenjoHoj,AleksColl}; 
and as will be evident in our formulas below, the RHS driving terms in 
the 2\textsuperscript{nd}-order wave equations can be made to vanish for 
propagation along a single axis. Alternatively, for the superpotential formalism 
of Kegeles and Cohen \cite{KegCohen} (upon which several other papers 
are based), the derivation is considerably involved; but it would appear that 
the homogeneous scalar wave equation (Eq.~\ref{ScalarWaves}) has been assumed 
by fiat, as has also been done -- appropriately in places, 
perhaps inappropriately in others -- in 
\cite{EspoScalarKas,AlhoBBwvEqn,OlPetKasner,OlPetThesis,CausticHartD}. 
(Interestingly, the discussion involving Equations $2.1-2.3$ of 
Alho {\it et al.} \cite{AlhoBBwvEqn} claims to show the scalar wave equation 
to be a sufficient condition for energy conservation, $\nabla^{a} T_{ab} = 0$; 
but no indication is given there to show the scalar wave equation to be 
a {\it necessary} condition for it.) Additionally, in Sagnotti and 
Zwiebach \cite{SagZwiebachWaves}, a subtle argument is made involving the 
introduction of time-dependent tetrad basis vectors in order to produce their 
Equation (1.18), a 2\textsuperscript{nd}-order wave equation derived for a 
Bianchi type-I background metric (of which the Kasner models are a subclass); 
this formula, from which all of the wave equations for their subsequent 
Kasner metric analyses are derived, is somehow missing the RHS driving term(s), 
and thus those terms are absent throughout the paper.

Given that studies such as these are analyzing many important aspects 
of electromagnetic wave propagation -- including field amplitude and energy bounds 
(i.e., whether or not field solutions ``blow up") close to the Kasner singularity 
\citep[e.g.,][]{OlPetKasner,AlhoBBwvEqn} -- it seems particularly important 
to avoid missing any terms which represent the exchange of energy between 
the different electric and magnetic field components (or, for that matter, 
between the electromagnetic and gravitational fields). And since this effect 
is due directly to the anisotropy of the expansion rates, it is reasonable to 
assume that such gravitationally-induced coupling of energy between the fields 
may very well occur in {\it any} anisotropic cosmological metric. It is therefore 
a real possibility that the evolution of electromagnetic energy in anisotropic 
cosmologies has been incompletely understood in all previous studies which 
have relied upon the homogeneous scalar wave equation for understanding 
vector fields such as light.

In this paper, the derivation of the wave equations for Kasner metrics with 
general indices will be carried out in Section~\ref{EMMath}, with some of 
the detailed steps shown in Appendix~\ref{App2ndOrderDeriv}. The complete 
solution of the fields for the $(p_{x}, p_{y}, p_{z}) = (1,0,0)$ case, 
for waves with general propagation directions, will be given in 
Section~\ref{SecKas100}, with some technical discussion of the link 
between this Kasner case and Minkowski spacetime given in 
Appendix~\ref{Kas100Mink}. 

Additionally, we note that while the Kasner wave equations are certainly 
apt for numerical calculations, the literature on this topic appears to be 
quite sparse. (Referring here specifically to numerical studies 
of the propagating electromagnetic fields on the predefined Kasner 
background metrics; not to numerical gravitational studies of the evolution 
of these metrics themselves, or of related spacetimes.) As a step towards 
providing results in this area, in Section~\ref{SecKas2323N13} we present the 
findings from a numerical simulation program that we have developed and applied 
to the study of the $\{p_{x}, p_{y}, p_{z}\} = \{(-1/3),(2/3),(2/3)\}$ case -- 
specifically to the propagating wave behavior of the electromagnetic field 
component along the axis of expansion rate symmetry. Notably, we demonstrate 
how the physical phase velocity of the light wave deviates from the 
null ray speed $(v = c \equiv 1)$, detailing how light propagation is slowed and 
ultimately stopped going back towards the singularity $(t \rightarrow 0)$ for this 
particular Kasner metric. Furthermore, we will recall an interesting class of 
cosmology-like metrics that reduce to this Kasner case as a vacuum limit, 
and will note how the formalism and numerical tools developed for this paper 
can be extended to study that class of metrics in future studies. 

Lastly, in Section~\ref{SecConclude}, our final discussion and summary 
of these results will be presented.

\section{Formulating the wave equations}
\label{EMMath}

\subsection{Electromagnetic conventions}
\label{EMconv}
For a metric $g_{\mu \nu}$ representing a spacetime 
containing only radiation (or vacuum), 
and no source charges or currents ($J^{\mu}=0$), 
the curved spacetime Maxwell equations in terms of the 
electromagnetic field tensor are \cite{MTW, WeinGC}: 
\begin{subequations}
\label{CurvMax:whole}
\begin{equation}
F_{\alpha \beta, \gamma} + F_{\beta \gamma, \alpha} 
+ F_{\gamma \alpha, \beta} = 0 ~,\label{CurvMax:a}
\end{equation}
\begin{equation}
F^{\alpha \beta}_{\phantom{000};\alpha} \propto 
(\sqrt{-|g_{\mu \nu}|} ~ F^{\alpha \beta})_{,\alpha}= 0 ~,\label{CurvMax:b}
\end{equation}
\end{subequations}
where $|g_{\mu \nu}|$ is the determinant of the metric tensor, 
semicolons refer to covariant derivatives, and commas to partial derivatives.

We wish to define the electric and magnetic fields in a way that will usefully 
represent the observable physical fields as seen by a ``fiducial" observer. 
In a metric with coordinates $(t,x,y,z)$, and a convenient comoving reference frame, 
this would represent a stationary observer with $dx = dy = dz = 0$ for all time.

For metrics with some symmetry-breaking physical behavior (such a distinct ``flow" 
of radiation in a particular direction), it might also be interesting to define 
a different class of fiducial observers (i.e., those stationary with respect 
to that flow); but for now, we will define our fields by considering a network of 
observers who are at least {\it instantaneously} stationary in $(x,y,z)$ position. 

Such observers will have timelike worldlines as long as $t$ is a timelike coordinate, 
with $g_{tt} < 0$. For the formalism presented here, the only assumptions we make are 
that $t$ is indeed timelike, and that (for considerable simplification) all metrics 
under consideration will be {\it diagonal}. We do not assume them to be synchronous or 
comoving, although all of the metrics used for calculations in this particular paper will be.

With those considerations, the normalized ($g_{\alpha \beta} U^{\alpha} U^{\beta} = -1$), 
future-directed timelike four-velocity of a fiducial observer at a 
given spacetime location will be:
\begin{equation}
U^{\alpha} = (1/\sqrt{-g_{tt}(t,x,y,z)}, 0, 0, 0) ~. \label{ObsFourVel}
\end{equation}

The electric and magnetic fields, respectively, measured by such an observer 
will be \cite{WaldGR}:
\begin{subequations}
\label{WaldFields:EM}
\begin{equation}
E_{\alpha} = F_{\alpha \beta} U^{\beta} ~,\label{WaldFields:E}
\end{equation}
\begin{equation}
B_{\alpha} = -\frac{1}{2} \epsilon _{\alpha \beta \gamma \delta} 
F^{\gamma \delta} U^{\beta} ~, \label{WaldFields:M}
\end{equation}
\end{subequations}
where $\epsilon _{\alpha \beta \gamma \delta}$ is the covariant 
Levi-Civita totally antisymmetric tensor.

For ease of calculations, we express the Levi-Civita tensor 
in terms of the Levi-Civita {\it symbol}, 
$\tilde{\epsilon} _{\alpha \beta \gamma \delta}$, a tensor density  
of weight $w=1$, filled solely with values $(+1,-1,0)$ \cite{CarrollGR}. 
Thus, for a metric that is diagonal:
\begin{equation}
\epsilon _{\alpha \beta \gamma \delta} = 
\sqrt{-|g_{\mu \nu}|} ~ \tilde{\epsilon} _{\alpha \beta \gamma \delta} = 
\sqrt{-g_{tt} g_{xx} g_{yy} g_{zz}} ~ \tilde{\epsilon} _{\alpha \beta \gamma \delta} ~. 
\label{Levi-Civ}
\end{equation}

Using the sign convention from \cite{WaldGR} 
of $\tilde{\epsilon} _{0123} \equiv \tilde{\epsilon} _{txyz} = +1$, 
each of Equations~\ref{WaldFields:M} can individually be inverted to yield 
the components of $F^{\gamma \delta}$ in terms of the $B$-fields, 
and then we can obtain the covariant version of the electromagnetic field tensor 
via $F_{\alpha \beta} = g_{\alpha \gamma} g_{\beta \delta}  F^{\gamma \delta}$. 
Similarly (and more simply), we can invert Equations~\ref{WaldFields:E} to 
yield: $F_{i0} = - F_{0i} = \sqrt{-g_{tt}} E_{i}$, for $i \ne 0$. So to sum up, 
for the sake of clarity -- given variations among different authors on the 
distribution of signs and metric factors -- we explicitly write out this tensor as:
\begin{equation}
F_{\alpha \beta} = 
\begin{bmatrix}
0 & -\sqrt{-g_{tt}} E_{x} & -\sqrt{-g_{tt}} E_{y} & -\sqrt{-g_{tt}} E_{z} \\
\sqrt{-g_{tt}} E_{x} & 0 & \sqrt{\frac{g_{xx} g_{yy}}{g_{zz}}} B_{z} 
& -\sqrt{\frac{g_{xx} g_{zz}}{g_{yy}}} B_{y} \\
\sqrt{-g_{tt}} E_{y} & -\sqrt{\frac{g_{xx} g_{yy}}{g_{zz}}} B_{z} 
& 0 & \sqrt{\frac{g_{yy} g_{zz}}{g_{xx}}} B_{x} \\
\sqrt{-g_{tt}} E_{z} & \sqrt{\frac{g_{xx} g_{zz}}{g_{yy}}} B_{y} 
& -\sqrt{\frac{g_{yy} g_{zz}}{g_{xx}}} B_{x} & 0 \\
\end{bmatrix} ~.\label{CovFabTensor}
\end{equation}
Employing these conventions for the electromagnetic tensor in 
Equations~\ref{CurvMax:whole} will produce observationally sensible 
definitions for the $E-$ and $B$-fields.

\subsection{Deriving wave equations for the evolving metric}
\label{KasMet}
We must now choose a particular class of metrics for the 
derivation of the curved spacetime Maxwell equations 
via Eq's.~\ref{CurvMax:whole},\ref{CovFabTensor}. Here we are interested 
in the well known Kasner metrics, the homogeneous but anisotropic spacetimes 
discussed above in the Introduction. They are as defined in Eq.~\ref{KasMetric}, 
where vacuum spacetimes are obtained by imposing the usual Kasner index conditions, 
specified by Eq.~\ref{KasParams}. 

There is no necessity to assume these conditions, however -- for example, the 
isotropic Friedmann universes filled with matter or radiation can be recovered 
via the choices (respectively) of $p_{x} = p_{y} = p_{z} = 2/3$ or $1/2$. 
We will place no a priori restrictions (other than realness) on the 
Kasner indices; we simply note that there must be some physical motivation 
for the nature of the cosmic mass-energy that results from a particular choice 
of $(p_{x}, p_{y}, p_{z})$, especially if such a choice violates 
reasonable energy conditions.

For our work here, we will be considering the application of these wave equations 
under the commonly used ``test field" approximation, in which we assume that 
the amplitudes and energy densities of the electromagnetic fields being studied 
are always small enough so that their gravitational perturbations onto the 
background metric can be neglected. (In cases where gravitational energy from the 
background metric can pump energy into the fields to increase their amplitudes 
significantly, it is therefore imperative to make sure that the test fields 
actually {\it stay} small enough during the propagation to continue neglecting 
their gravitational effects.)

Applying the formulas of Section~\ref{EMconv} to this metric, we see that 
the electromagnetic field tensor becomes:
\begin{equation}
F_{\alpha \beta} = 
\begin{bmatrix}
0 & -E_{x} & -E_{y} & -E_{z} \\
E_{x} & 0 & t^{(p_{x} + p_{y} - p_{z})} B_{z} 
& -t^{(p_{x} - p_{y} + p_{z})} B_{y} \\
E_{y} & -t^{(p_{x} + p_{y} - p_{z})} B_{z} 
& 0 & t^{(-p_{x} + p_{y} + p_{z})} B_{x} \\
E_{z} & t^{(p_{x} - p_{y} + p_{z})} B_{y} 
& -t^{(-p_{x} + p_{y} + p_{z})} B_{x} & 0 \\
\end{bmatrix} ~.\label{KasCovEMTensor}
\end{equation}

Equations~\ref{CurvMax:whole} then give us eight distinct field equations.
Two of these equations (one each from Eq.~\ref{CurvMax:a} and Eq.~\ref{CurvMax:b}) 
are modified versions of the usual divergence equations; dividing out common 
powers of $t$, we can write them as:
\begin{subequations}
\label{DivergenceEqs}
\begin{equation}
t^{-2 p_{x}} E_{x,\phantom{.} x} + t^{-2 p_{y}} E_{y,\phantom{.} y} 
+ t^{-2 p_{z}} E_{z,\phantom{.} z} = 0 ~,\label{EDivergence}
\end{equation}
\begin{equation}
t^{-2 p_{x}} B_{x,\phantom{.} x} + t^{-2 p_{y}} B_{y,\phantom{.} y} 
+ t^{-2 p_{z}} B_{z,\phantom{.} z} = 0 ~. \label{BDivergence}
\end{equation}
\end{subequations}

The remaining six equations are the ``curl"-like equations, which are:
\begin{subequations}
\label{CurlEqs}
\begin{equation}
(B_{z,\phantom{.} y} - B_{y,\phantom{.} z}) 
= [t^{(-p_{x} + p_{y} + p_{z})} E_{x}]_{,\phantom{.} t} ~, \label{ExCurl}
\end{equation}
\begin{equation}
(E_{z,\phantom{.} y} - E_{y,\phantom{.} z}) 
= - [t^{(-p_{x} + p_{y} + p_{z})} B_{x}]_{,\phantom{.} t} ~, \label{BxCurl}
\end{equation}
\end{subequations}
plus four more analogous equations obtained through cyclic permutations 
over $(x,y,z)$ and corresponding factors indexed to $(x,y,z)$ -- 
such as $(p_{x},p_{y},p_{z})$, and spatial partial derivatives.

These strongly resemble the usual flat spacetime Maxwell equations, 
just with the appearance of additional powers of $t$; and at first glance, 
it might seem possible to absorb these factors through alternative definitions 
of the fields comprising $F_{\alpha \beta}$. For example, with the redefinitions:
\begin{equation}
E^{'}_{x} \equiv t^{(-p_{x} + p_{y} + p_{z})} E_{x} ~, ~ 
B^{'}_{x} \equiv t^{(-p_{x} + p_{y} + p_{z})} B_{x} ~,
\label{PrimeFieldsReDef}
\end{equation}
(and cyclic permutations thereof), the divergence equations (Eq's.~\ref{DivergenceEqs}) 
can be simplified to $\nabla \cdot \bm{E}^{'} = \nabla \cdot \bm{B}^{'} = 0$ 
(using the flat spacetime version of the dot product here), and the right hand sides 
of Eq's.~\ref{CurlEqs} reduce simply to the relevant spatial components of 
$\bm{E}^{'}_{,\phantom{.} t}$, $\bm{B}^{'}_{,\phantom{.} t}$. However, 
additional factors appear on the left hand sides of the equations as a consequence; 
for example, Eq's.~\ref{CurlEqs} become:
\begin{subequations}
\label{CurlReDefs}
\begin{equation}
[t^{(-p_{x} - p_{y} + p_{z})} B^{'}_{z,\phantom{.} y}] 
- [t^{(-p_{x} + p_{y} - p_{z})} B^{'}_{y,\phantom{.} z}]
= E^{'}_{x,\phantom{.} t} ~, \label{ExCurlReDef}
\end{equation}
\begin{equation}
[t^{(-p_{x} - p_{y} + p_{z})} E^{'}_{z,\phantom{.} y}] 
- [t^{(-p_{x} + p_{y} - p_{z})} E^{'}_{y,\phantom{.} z}] 
= - B^{'}_{x,\phantom{.} t} ~, \label{BxCurlReDef}
\end{equation}
\end{subequations}
(and cyclic permutations), which does not succeed in removing the 
(physically meaningful) powers of $t$ in these equations, but 
just shifts them around. However, these redefinitions do succeed 
in greatly simplifying the derivations and forms of the wave equations 
(as well as their eventual solutions), so we will work with these 
redefined fields from this point on.

Normally, the next step is to convert the coupled 1\textsuperscript{st}-order 
differential equations for the fields into uncoupled 2\textsuperscript{nd}-order 
wave equations. In charge-free Minkowski space, where the fields obey 
$\nabla \cdot \bm{E} = \nabla \cdot \bm{B} = 0$, 
$\nabla \times \bm{E} = -\bm{B}_{,\phantom{.} t}$, and 
$\nabla \times \bm{B} = \bm{E}_{,\phantom{.} t}$, the usual trick is to apply vector 
identities to write $\nabla \times \nabla \times \bm{E} = 
\nabla (\nabla \cdot \bm{E}) - \nabla ^{2} \bm{E} = - \nabla ^{2} \bm{E}$, and then use 
$\nabla \times \nabla \times \bm{E} = - (\nabla \times \bm{B}) _{,\phantom{.} t} 
= -\bm{E}_{,\phantom{.} t,\phantom{.} t}$ 
to complete the wave equation. But in the Kasner case, there are 
unavoidable powers of $t$ in Eq's.~\ref{CurlReDefs} which do not commute 
with the time derivatives $\partial / \partial t$ during the aforementioned steps. 
Therefore, the resulting 2\textsuperscript{nd}-order differential equations 
are {\it not} entirely uncoupled here, in general. 

Applying the same ``curl of a curl" trick with these equations anyway, 
modified as necessary by these extra powers of $t$, after some work we derive:
\begin{eqnarray}
 E^{'}_{x,\phantom{.} t,\phantom{.} t} 
+ \left(\frac{p_{x}}{t}\right) E^{'}_{x,\phantom{.} t} 
- [(t^{-2 p_{x}} E^{'}_{x,\phantom{.} x,\phantom{.} x}) 
+ (t^{-2 p_{y}} E^{'}_{x,\phantom{.} y,\phantom{.} y}) 
+ (t^{-2 p_{z}} E^{'}_{x,\phantom{.} z,\phantom{.} z})] = \nonumber \\ 
\frac{(p_{z} - p_{y})}{t} \{ [t^{(-p_{x} - p_{y} + p_{z})} B^{'}_{z,\phantom{.} y}] 
+ [t^{(-p_{x} + p_{y} - p_{z})} B^{'}_{y,\phantom{.} z}] \} ~, \label{GenKas2ndOrdEx}
\end{eqnarray}
where similar equations for $E^{'}_{y}$ and $E^{'}_{z}$ can be obtained 
from this via cyclic permutations over $(x,y,z)$ and corresponding factors; 
and the analogous equations for the magnetic fields are obtained via the 
substitutions $E^{'}_{i} \rightarrow B^{'}_{i}$, $B^{'}_{i} \rightarrow -E^{'}_{i}$.
The full details of this derivation are given in Appendix~\ref{App2ndOrderDeriv}.

Note first that the {\it plus} sign between the two terms on the right hand side (RHS) 
of Equation~\ref{GenKas2ndOrdEx}, instead of a minus sign, means that the RHS cannot be 
eliminated via any manipulations of the curl-like equations, Eq's.~\ref{CurlReDefs}. 
(These are, in fact, the field-coupling nonhomogeneous terms discussed significantly 
in the Introduction.) Using Eq's.~\ref{CurlReDefs} in conjunction with Eq.~\ref{GenKas2ndOrdEx}, 
however, does allow us to present this wave equation in multiple alternative forms. 
Using the shorthand notation 
of $``\{t \nabla ^{2}\}" \equiv [(t^{-2 p_{x}}) \partial^{2}/\partial x ^{2} 
+ (t^{-2 p_{y}}) \partial^{2}/\partial y ^{2} 
+ (t^{-2 p_{z}}) \partial^{2}/\partial z ^{2}]$, we can present the 
2\textsuperscript{nd}-order wave equation for $E^{'}_{x}$ (and analogously 
for the other fields) in three useful variants for each field: 
\begin{subequations}
\label{ExWaveEqn3ways}
\begin{equation}
 E^{'}_{x,\phantom{.} t,\phantom{.} t} 
+ \left(\frac{p_{x}}{t}\right) E^{'}_{x,\phantom{.} t} 
- \{t \nabla ^{2}\} E^{'}_{x} = 
\frac{(p_{z} - p_{y})}{t} \{ [t^{(-p_{x} - p_{y} + p_{z})} B^{'}_{z,\phantom{.} y}] 
+ [t^{(-p_{x} + p_{y} - p_{z})} B^{'}_{y,\phantom{.} z}] \} ~,~ or, \label{ExWaveEqnWay1}
\end{equation}
\begin{equation}
E^{'}_{x,\phantom{.} t,\phantom{.} t} 
+ \left[\frac{p_{x} + (p_{z}-p_{y})}{t}\right] E^{'}_{x,\phantom{.} t} 
- \{t \nabla ^{2}\} E^{'}_{x} = 2 \frac{(p_{z} - p_{y})}{t} 
\{ [t^{(-p_{x} - p_{y} + p_{z})} B^{'}_{z,\phantom{.} y}] \} ~,~ 
or, \phantom{0000000000} \label{ExWaveEqnWay2}
\end{equation}
\begin{equation}
E^{'}_{x,\phantom{.} t,\phantom{.} t} 
+ \left[\frac{p_{x} - (p_{z}-p_{y})}{t}\right] E^{'}_{x,\phantom{.} t} 
- \{t \nabla ^{2}\} E^{'}_{x} = 2 \frac{(p_{z} - p_{y})}{t} 
\{ [t^{(-p_{x} + p_{y} - p_{z})} B^{'}_{y,\phantom{.} z}] \} 
~. \phantom{000000000} \label{ExWaveEqnWay3}
\end{equation}
\end{subequations}
Rather than creating ambiguity, however, having these different variants available 
will end up aiding us in deriving further wave equations that are ultimately uncoupled. 

These equations take the form of damped (or pumped), driven oscillators. 
The nonhomogeneous driving terms on the RHS of these 
2\textsuperscript{nd}-order wave equations can be interpreted as the 
anisotropically evolving spacetime behaving somewhat like an active medium, 
causing the magnetic fields to drive the electric fields, and vice versa, in a 
more involved manner than happens in Minkowski space \citep[e.g.,][]{TsagasCurvEM}. 

In a specific set of cases, the RHS terms do properly equal zero -- particularly 
in the case of axisymmetric Kasner metrics (e.g., $p_{y} = p_{z}$), for the fields 
along the distinctly-evolving direction (e.g., $E_{x}$ and $B_{x}$). But for 
other fields and cases, dropping the RHS terms causes one to lose about half 
of the solutions, as we will see below.

Examining these 2\textsuperscript{nd}-order partial differential equations, 
a way to simplify them immediately is through separation of variables. For each 
electric or magnetic field, $F^{'}_{i}$ assume the following solution: 
\begin{eqnarray}
& & F^{'}_{i} (t,x,y,z) = F^{''}_{i} (t) ~ S_{i} (x,y,z) = \nonumber \\ 
& & = F^{''}_{i} (t) [c_{1} \sin (k_{x} x) + c_{2} \cos (k_{x} x)] 
[c_{3} \sin (k_{y} y) + c_{4} \cos (k_{y} y)] 
[c_{5} \sin (k_{z} z) + c_{6} \cos (k_{z} z)] ~, \label{SepVars}
\end{eqnarray}
where $ \{c_{1},\dots,c_{6} \} $ are arbitrary constants dependent upon 
initial conditions. (To simplify our numerical calculations, we have 
chosen to work with real functions throughout.) The constants 
$(k_{x}, k_{y}, k_{z}) \equiv \bm{k}$ are real, positive numbers 
with $k \equiv \sqrt{k_{x}^{2} + k_{y}^{2} + k_{z}^{2}}$. 
This form of solution means that the light waves ``breathe" with the expansion 
or contraction along different axis directions, as a particular set of 
coordinate distances, say $(\Delta x, \Delta y, \Delta z)$, represents different 
physical distances at different times. 

With substitutions like Eq.~\ref{SepVars}, Equation~\ref{GenKas2ndOrdEx} becomes: 
\begin{eqnarray}
E^{''}_{x,\phantom{.} t,\phantom{.} t} 
+ \left(\frac{p_{x}}{t}\right) E^{''}_{x,\phantom{.} t} 
+ \left(\frac{k_{x}^{2}}{t^{2 p_{x}}} + \frac{k_{y}^{2}}{t^{2 p_{y}}} 
+ \frac{k_{z}^{2}}{t^{2 p_{z}}}\right) E^{''}_{x} 
= \frac{1}{S_{x} (x,y,z)} \times \nonumber \\ 
\frac{(p_{z} - p_{y})}{t} \{ [t^{(-p_{x} - p_{y} + p_{z})} B^{'}_{z,\phantom{.} y}] 
+ [t^{(-p_{x} + p_{y} - p_{z})} B^{'}_{y,\phantom{.} z}] \} ~, 
\label{GenKas2ndOrdkxkykz}
\end{eqnarray}
where equations for all of the other fields are obtained from this 
via cyclic permutations as before, and where the alternative versions 
for each field, as given in Eqn's.~\ref{ExWaveEqn3ways}, still apply.

Now, it should be understood that the 
Equations~\ref{GenKas2ndOrdEx}-\ref{GenKas2ndOrdkxkykz} were written assuming that 
$(k_{x}, k_{y}, k_{z})$ are nonzero. Letting one or two of these wavenumbers equal zero 
(e.g., propagation along a Kasner axis) could cause confusion. The form of the 
spatial functions $S_{i} (x,y,z)$, 
plus $\nabla \cdot \bm{E}^{'} = \nabla \cdot \bm{B}^{'} = 0$, clearly shows 
that the wave fields are transverse to the instantaneous propagation direction. 
So for example, suppose $\bm{k} = (k_{x}, 0, 0)$. Clearly all of the fields then 
satisfy $F^{'}_{i,\phantom{.} y} = F^{'}_{i,\phantom{.} z} = 0$, and 
Equations~\ref{ExWaveEqn3ways} seem to imply that ``adding zero" to the RHS 
in different ways gives us three meaningfully different versions of the left hand side. 
But in this case, the transverse nature of the fields tells us that $E^{'}_{x} = 0$ 
(as well as $B^{'}_{x} = 0$), hence all three versions are moot. Alternatively, if we suppose 
$\bm{k} = (0, k_{y}, 0)$, then all three versions make sense; with the simplest one 
being Eq.~\ref{ExWaveEqnWay3}, where the RHS now does equal zero, and the coefficient 
of the $(E^{'}_{x,\phantom{.} t}/t)$ term is $[p_{x} - (p_{z}-p_{y})]$. If one assumes 
the usual Kasner condition, $(p_{x} + p_{y} + p_{z}) = 1$, then our resulting equation 
is equivalent to Equation (3.18) of Sagnotti and Zwiebach \cite{SagZwiebachWaves} 
for this special case.

To ultimately disentangle the wave equations, we can use the alternative formulations 
of Eq's.~\ref{ExWaveEqn3ways} to tailor the couplings between the electric and 
magnetic fields to group them as three disconnected pairs of fields -- for example, 
$(E_{x} \leftrightarrow B_{y})$, $(E_{y} \leftrightarrow B_{z})$, and 
$(E_{z} \leftrightarrow B_{x})$. The equations for the first pair would look like this:
\begin{subequations}
\label{CoupledPair2ndOrd}
\begin{equation}
E^{'}_{x,\phantom{.} t,\phantom{.} t} 
+ \left[\frac{p_{x} - (p_{z}-p_{y})}{t}\right] E^{'}_{x,\phantom{.} t} 
- \{t \nabla ^{2}\} E^{'}_{x} = 2 \frac{(p_{z} - p_{y})}{t} 
\{ [t^{(-p_{x} + p_{y} - p_{z})} B^{'}_{y,\phantom{.} z}] \} ~, \label{CoupledPairEx}
\end{equation}
\begin{equation}
B^{'}_{y,\phantom{.} t,\phantom{.} t} 
+ \left[\frac{p_{y} + (p_{x}-p_{z})}{t}\right] B^{'}_{y,\phantom{.} t} 
- \{t \nabla ^{2}\} B^{'}_{y} = -2 \frac{(p_{x} - p_{z})}{t} 
\{ [t^{(p_{x} - p_{y} - p_{z})} E^{'}_{x,\phantom{.} z}] \} ~, \label{CoupledPairBy}
\end{equation}
\end{subequations}

We then take $\partial / \partial z$ of Eq.~\ref{CoupledPairEx}, apply our spatial functions 
from Eq.~\ref{SepVars} as necessary, and invert to get:
\begin{eqnarray}
& & B^{'}_{y} = -\left(\frac{1}{k_{z}^{2}}\right) \left[\frac{t}{2 (p_{z} - p_{y})}\right] 
[t^{(p_{x} - p_{y} + p_{z})}] \times \nonumber \\
& & \left\{ E^{'}_{x,\phantom{.} t,\phantom{.} t, \phantom{.} z} 
+ \left[\frac{p_{x} 
- (p_{z}-p_{y})}{t}\right] E^{'}_{x,\phantom{.} t, \phantom{.} z} 
+ \left(\frac{k_{x}^{2}}{t^{2 p_{x}}} + \frac{k_{y}^{2}}{t^{2 p_{y}}} 
+ \frac{k_{z}^{2}}{t^{2 p_{z}}}\right) E^{'}_{x, \phantom{.} z} \right\}
 ~, \label{CoupledExProcessed}
\end{eqnarray}

This expression for $B^{'}_{y}$ is then inserted back into Eq.~\ref{CoupledPairBy} 
to eliminate it in favor of $E^{'}_{X}$. Then we apply $\partial / \partial z$ once 
more -- getting another common factor of $(-k_{z}^{2})$ -- and then divide out the 
spatial functions, and any other factors in front of the highest time derivative term. 
The result is a {\it fourth}-order ordinary differential equation for 
$E^{''}_{x} (t)$. This equation is the {\it same} for the magnetic fields also 
(no sign changes or other factors). So for $F^{''}_{x} (t) \equiv E^{''}_{x} (t)$ 
or $B^{''}_{x} (t)$, we have the formula:
\begin{widetext}
\begin{eqnarray}
& & F^{''}_{x,\phantom{.} t,\phantom{.} t,\phantom{.} t,\phantom{.} t} 
+ \left[\frac{2 (1 + 2 p_{x})}{t}\right] F^{''}_{x,\phantom{.} t,\phantom{.} t,\phantom{.} t} 
\nonumber \\
& & + \left\{2 \left( \frac{k_{x}^{2}}{t^{2 p_{x}}} + \frac{k_{y}^{2}}{t^{2 p_{y}}} 
+ \frac{k_{z}^{2}}{t^{2 p_{z}}} \right) 
+ \frac{1}{t^{2}} [2 p_{x} + 5 p_{x}^{2} - (p_{y} - p_{z})^{2}] \right\} 
F^{''}_{x,\phantom{.} t,\phantom{.} t} 
\nonumber \\
& & + \boldsymbol{\left(}\frac{2}{t} \left\{ \frac{k_{x}^{2}}{t^{2 p_{x}}}
+ [1 + 2 (p_{x} - p_{y})] \frac{k_{y}^{2}}{t^{2 p_{y}}} 
+ [1 + 2 (p_{x} - p_{z})] \frac{k_{z}^{2}}{t^{2 p_{z}}} \right\}
+ \{ \frac{1}{t^{3}} (2 p_{x} - 1) [p_{x}^{2} - (p_{y} - p_{z})^{2}] \} \boldsymbol{\right)}
F^{''}_{x,\phantom{.} t} 
\nonumber \\
& & + \boldsymbol{\left[} \left( \frac{k_{x}^{2}}{t^{2 p_{x}}} 
+ \frac{k_{y}^{2}}{t^{2 p_{y}}} 
+ \frac{k_{z}^{2}}{t^{2 p_{z}}} \right)^{2}
+ \frac{2}{t^{2}} 
\{ [(p_{x} - p_{y}) (1 + p_{x} + p_{z} - 3 p_{y}) \frac{k_{y}^{2}}{t^{2 p_{y}}}] 
+ [(p_{x} - p_{z}) (1 + p_{x} + p_{y} - 3 p_{z}) \frac{k_{z}^{2}}{t^{2 p_{z}}}] \} 
\boldsymbol{\right]}
\nonumber \\
& & \times F^{''}_{x} = 0 ~, \label{GenKas4thOrderDiffEq}
\end{eqnarray} 
\end{widetext}
with the analogous equations for $\{ E^{''}_{y}, B^{''}_{y} \}$, $\{E^{''}_{z}, B^{''}_{z} \}$ 
obtained as usual through cyclic permutations over $(x,y,z)$, $(p_{x},p_{y},p_{z})$, 
and $(k_{x},k_{y},k_{z})$. (Note that we get the exact same 
4\textsuperscript{th}-order equations even if we derive them 
via $(E_{x} \leftrightarrow B_{z})$ instead of $(E_{x} \leftrightarrow B_{y})$, 
and so on.)

The above procedure is only valid for fields where the RHS of formulas 
like Eq.~\ref{CoupledPairEx} are nonzero. For those where the RHS is zero -- 
say, through $k_{y} = k_{z} = 0$ (in which case the field itself is zero), 
or through $p_{y} = p_{z}$ (axisymmetric Kasner case), where the RHS for 
the $F^{''}_{x}$ fields (but {\it not} for the $F^{''}_{y}$ or $F^{''}_{z}$ 
fields) are zero -- then the 2\textsuperscript{nd}-order wave equation is all 
we have for that field, and we only have two (if any) nontrivial solutions 
for it. But, for the general case in which Equation~\ref{GenKas4thOrderDiffEq} 
holds, then we actually have four independent solutions for the 
temporal wave function. In the next section, we will see that the correct 
set of solutions to use depends upon the {\it polarization} of the waves.

While Eq.~\ref{GenKas4thOrderDiffEq} appears well suited for numerical 
calculations, one might despair of finding analytical solutions (or any 
clear understanding of the solutions) for a 4\textsuperscript{th}-order 
differential equation, especially one as complicated as this one. 
But that is not necessarily the case.

First, just from the nature of the 2\textsuperscript{nd}-order 
equations, it is clear that the ubiquitous expression 
$(t^{-2 p_{x}} k_{x}^{2} + t^{-2 p_{y}} k_{y}^{2} 
+ t^{-2 p_{z}} k_{z}^{2})$ tells us a great deal about the 
effective frequency of oscillations at all times, especially in early-, mid-, 
or late-time regimes where one or another of the three factors is dominant. 
Also, the presence of the term $(F^{''}_{i,\phantom{.} t}/t)$ implies 
that we will have Bessel-function-like behavior as $t \rightarrow 0$, 
but almost pure sinusoidal behavior (with adiabatically-varying frequency) 
at late times. Much of the qualitative behavior of the fields is therefore obvious.

Furthermore, we will show that it is also possible to ``guess" the exact solution 
to the 4\textsuperscript{th}-order equation, by inferring it from known 
solutions to a 2\textsuperscript{nd}-order equation like Eq.~\ref{GenKas2ndOrdkxkykz} 
when the RHS happens to equal zero. (In fact, it may not be a bad trick to try this 
in general, by temporarily setting the RHS to zero to search for solutions, 
even if the RHS does not properly equal zero for that case.) It is an 
unfortunate drawback that even for 2\textsuperscript{nd}-order equations 
of this type with a RHS equal to zero, known solutions still seem to be rare; 
but this is a promising path to follow when possible.

For axisymmetric Kasner models, for which two of the parameters $p_{i}$ 
are equal, the RHS of the 2\textsuperscript{nd}-order equation for the 
fields parallel to the axis of symmetry is automatically equal to zero. 
If one sticks to the usual parameter constraints of Eq.~\ref{KasParams}, 
then the only two possibilities are the models: 
$(p_{1}, p_{2}, p_{3}) = (1,0,0)$, and 
$\{p_{1}, p_{2}, p_{3}\} = \{(-1/3),(2/3),(2/3)\}$. These are therefore 
the most heavily studied Kasner cases. Although our formalism so far has 
placed no such restrictions on the $p_{i}$ parameters, we will nevertheless 
focus the rest of this paper upon these two cases, because of these reasons: 
the first case is (almost uniquely) completely solvable for all of the fields, 
as we will show; the second case exhibits interesting violations of 
geometric optics as $t \rightarrow 0$; and, the second case is the 
vacuum limit of a class of non-vacuum, inhomogeneous metrics of 
particular interest to us for future cosmological study.

\section{Kasner special case $\bm{(1,0,0)}$}
\label{SecKas100}
Choosing the axes such that $(p_{x}, p_{y}, p_{z}) = (1,0,0)$, 
the metric becomes:
\begin{equation}
ds^{2} = -dt^{2} + t^{2} dx^{2} + dy^{2} + dz^{2} ~. 
\label{Kas100Metric}
\end{equation}

Now, the parameter choices $\bm{p} = (1,0,0)$ (or equivalently 
$(0,1,0)$ or $(0,0,1)$) represents a unique Kasner case because it is 
actually a flat spacetime -- all Riemann tensor components are zero. 
It can therefore be transformed into the Minkowski metric, with the 
implication that the field equations must be those of ordinary 
flat spacetime, with their purely sinusoidal solutions. 
But the formulas to be presented in this section will 
not {\it look} very much like those of flat spacetime.

This is due to the fact that the transformation producing the metric 
in Equation~\ref{Kas100Metric} has the effect of nontrivially mixing 
together the different coordinates and field components. In particular, 
note that the ``separation of variables" from Eq.~\ref{SepVars} -- 
with terms like $\sin (k_{x} x)$ -- is not at all the same kind of 
separation of variables that one would do in static Minkowski spacetime, 
since the argument ``$k_{x} x$" actually represents a physical distance 
that increases in time like $t$, rather than a static wavelength. 

It is therefore important to remember that the solutions presented here 
for this case are merely an unusual way of combining and transforming 
the different flat spacetime solutions. (A derivation of how 
Kasner $(1,0,0)$ fields are related to the Minkowski frame fields 
is given in Appendix~\ref{Kas100Mink}.) Nevertheless, it is instructive 
to use this metric to demonstrate the method of finding solutions, 
and their properties; and it is probably not a coincidence that the 
least physically complicated case is also the most mathematically solvable one.

Referring back to Eq.~\ref{GenKas2ndOrdkxkykz}, the wave equation 
for $F^{''}_{x} (t) \in \{E^{''}_{x} (t), B^{''}_{x} (t)\}$ now becomes:
\begin{equation}
F^{''}_{x,\phantom{.} t,\phantom{.} t} + \left( \frac{1}{t} \right) F^{''}_{x,\phantom{.} t} 
+ \left( \frac{k_{x}^{2}}{t^{2}} + k_{y}^{2} 
+ k_{z}^{2} \right) F^{''}_{x} = 0 ~, \label{Kas1002ndOrdExBx}
\end{equation}
which we recognize as an example of the transformed Bessel equation \cite{CRCmath} 
with solutions $J_{\pm i k_{x}}[(k_{y}^{2} + k_{z}^{2})^{1/2} t]$. For 
Bessel functions of purely imaginary order like these, real solutions 
(for $t > 0$) can be constructed \cite{DunsterImagBessel} as:
\begin{eqnarray}
J^{2+}_{i k_{x}}[(k_{y}^{2} + k_{z}^{2})^{1/2} t] \equiv 
\frac{1}{2} \{ J_{i k_{x}}[(k_{y}^{2} + k_{z}^{2})^{1/2} t] 
+ J_{-i k_{x}}[(k_{y}^{2} + k_{z}^{2})^{1/2} t] \} ~, \nonumber\\ 
J^{2-}_{i k_{x}}[(k_{y}^{2} + k_{z}^{2})^{1/2} t] \equiv 
\frac{1}{2i} \{ J_{i k_{x}}[(k_{y}^{2} + k_{z}^{2})^{1/2} t] 
- J_{-i k_{x}}[(k_{y}^{2} + k_{z}^{2})^{1/2} t] \} ~. 
\label{Kas1002ndOrdRealSolns}
\end{eqnarray}
These are equivalent to (i.e., linear combinations of) the 
previously known solutions, as given in (for example) 
Sagnotti and Zwiebach \cite{SagZwiebachWaves} and 
Petersen \cite{OlPetKasner}. They turn out to represent 
only half of the space of solutions for this Kasner case, however. 

For the four remaining fields, the right hand sides of the 
2\textsuperscript{nd}-order wave equations cannot generally be made 
equal to zero; but those driving terms can be manipulated, 
analogously with Eq's.~\ref{ExWaveEqn3ways}, to make them dependent only 
upon $E^{'}_{x} (t)$ or $B^{'}_{x} (t)$. All of those equations 
end up having the same form: 
\begin{equation}
F^{'}_{a,\phantom{.} t,\phantom{.} t} + \left(\frac{1}{t}\right) F^{'}_{a,\phantom{.} t} 
+ \left( \frac{k_{x}^{2}}{t^{2}} + k_{y}^{2} + k_{z}^{2} \right) F^{'}_{a} 
= 2 F^{'}_{b} ~, \label{Kas100Coupled2ndOrd}
\end{equation}
with $(F^{'}_{a}, F^{'}_{b})$ representing the pairs of fields, respectively: 
$(E^{'}_{y}, B^{'}_{x,\phantom{.} z})$, $(E^{'}_{z}, -B^{'}_{x,\phantom{.} y})$, 
$(B^{'}_{y}, -E^{'}_{x,\phantom{.} z})$, and $(B^{'}_{z}, E^{'}_{x,\phantom{.} y})$. 

We note that this is the same differential equation (with the same temporal solutions) 
as Eq.~\ref{Kas1002ndOrdExBx} for $E^{''}_{x} (t)$ and $B^{''}_{x} (t)$, {\it if} 
we can set the right hand sides equal to zero. If not, then each field obeys the 
relevant 4th-order equation instead. A simplification of the physical situation 
therefore presents itself, where we break the solutions down into the two 
distinct, nontrivial possibilities: (i) Polarization ``$X_{E}$", where 
$E^{'}_{x} \neq 0$, $B^{'}_{x} = 0$, so that 
$(E^{''}_{x}, E^{''}_{y}, E^{''}_{z})$ are all linear combinations of the 
functions $(J^{2+}, J^{2-})$ from Eq's.~\ref{Kas1002ndOrdRealSolns}, 
and $(B^{''}_{y}, B^{''}_{z})$ are solutions of the 
4\textsuperscript{th}-order equations; and, (ii) Polarization ``$X_{B}$", 
where $E^{'}_{x} = 0$, $B^{'}_{x} \neq 0$, and vice-versa 
($E \leftrightarrow B$) for the solutions of the nonzero fields. 
The general solution can then be given as an appropriate combination of 
Polarization $X_{E}$ and Polarization $X_{B}$.

Now recalling Eq.~\ref{GenKas4thOrderDiffEq}, as applied to this 
Kasner $(1,0,0)$ metric, we obtain the 4th-order wave equation:
\begin{eqnarray}
& & F^{''}_{i,\phantom{.} t,\phantom{.} t,\phantom{.} t,\phantom{.} t} 
+ \left(\frac{2}{t}\right) F^{''}_{i,\phantom{.} t,\phantom{.} t,\phantom{.} t} 
+ \left[2 \left(\frac{k_{x}^{2}}{t^{2}} + k_{y}^{2} + k_{z}^{2} \right) 
- \frac{1}{t^{2}} \right] F^{''}_{i,\phantom{.} t,\phantom{.} t} 
\nonumber \\
& & + \left[\frac{2}{t} \left(-\frac{k_{x}^{2}}{t^{2}} + k_{y}^{2} + k_{z}^{2} \right) 
+ \frac{1}{t^{3}} \right] F^{''}_{i,\phantom{.} t} 
+ \left[ \left(\frac{k_{x}^{2}}{t^{2}} + k_{y}^{2} + k_{z}^{2} \right)^{2} 
+ \left( \frac{4}{t^{4}} k_{x}^{2} \right) \right] F^{''}_{i} 
= 0 ~, \label{Kas1004thOrderEqs}
\end{eqnarray}
which turns out to be the same equation applicable for all four remaining fields, 
$F^{''}_{i} \in \{E^{''}_{y} (t), B^{''}_{y} (t), E^{''}_{z} (t), B^{''}_{z} (t)\}$. 
One can immediately verify that two good solutions 
to this 4\textsuperscript{th}-order equation 
are $(J^{2+}, J^{2-})$ from Eq's.~\ref{Kas1002ndOrdRealSolns}; 
or in other words, $J_{i k_{x}}[(k_{y}^{2} + k_{z}^{2})^{1/2} t]$ 
and $J_{-i k_{x}}[(k_{y}^{2} + k_{z}^{2})^{1/2} t]$ are still good solutions. 
We must therefore find the two remaining ones. 

One interesting possibility for doubling a single Bessel function solution 
into two solutions, is by recalling their recurrence relations. These Bessel 
functions of imaginary order satisfy the usual relation \cite{DunsterImagBessel}:
\begin{equation}
J_{\pm i k_{x}}[(k_{y}^{2} + k_{z}^{2})^{1/2} t] = 
\left[ \frac{(k_{y}^{2} + k_{z}^{2})^{1/2}}{(\pm i k_{x})} \right]
\left( \frac{t}{2}  \right) 
\{ J_{(\pm i k_{x})-1}[(k_{y}^{2} + k_{z}^{2})^{1/2} t] \bm{+} 
J_{(\pm i k_{x})+1}[(k_{y}^{2} + k_{z}^{2})^{1/2} t] \} ~.
\label{BesselRecurPlus}
\end{equation}
Two interesting trial solutions can therefore be obtained by flipping 
the sign between the two contributing Bessel functions (i.e., the sign 
in bold above), and using the other recurrence relation involving the 
Bessel function time derivative: 
\begin{eqnarray}
& & \left[ (k_{y}^{2} + k_{z}^{2})^{1/2} \right]
\left( \frac{t}{2}  \right) 
\{ J_{(\pm i k_{x})-1}[(k_{y}^{2} + k_{z}^{2})^{1/2} t] \bm{-} 
J_{(\pm i k_{x})+1}[(k_{y}^{2} + k_{z}^{2})^{1/2} t] \} 
\nonumber \\ 
& & = t \{ J_{\pm i k_{x}}[(k_{y}^{2} + k_{z}^{2})^{1/2} t]_{, \phantom{.} t} \} ~.
\label{BesselRecurMinus}
\end{eqnarray}
Explicit substitution into Eq.~\ref{Kas1004thOrderEqs} does indeed verify that 
$\bm{(}t \{ J_{\pm i k_{x}}[(k_{y}^{2} + k_{z}^{2})^{1/2} t]_{, \phantom{.} t} \}\bm{)}$ 
are indeed the two remaining solutions. To this author's knowledge, these 
expressions have not been identified elsewhere by other researchers as solutions 
to the wave equations in the Kasner $(1,0,0)$ metric. 

One way of organizing the basis set of four independent solutions could be 
by using the four variations, 
$\bm{(} t \{ J_{(\pm i k_{x}) \pm 1}[(k_{y}^{2} + k_{z}^{2})^{1/2} t] \} \bm{)}$. 
In practice, however, our choice is to organize them into purely real solutions, 
which we do as follows. Let $\omega_{yz} \equiv (k_{y}^{2} + k_{z}^{2})^{1/2}$, 
and $k_{R} \equiv (\omega_{yz}/k_{x})$. Then, using the recurrence relations in 
Eq's.~\ref{BesselRecurPlus}-\ref{BesselRecurMinus}, the previously defined 
solutions $(J^{2+}, J^{2-})$ from Eq's.~\ref{Kas1002ndOrdRealSolns} become:
\begin{eqnarray}
J^{2+}_{i k_{x}}(\omega_{yz} t) = \left(\frac{k_{R} \phantom{.} t}{4 i}\right) 
\{ J_{(i k_{x})-1}(\omega_{yz} t) + J_{(i k_{x})+1}(\omega_{yz} t) 
- J_{(-i k_{x})-1}(\omega_{yz} t) - J_{(-i k_{x})+1}(\omega_{yz} t) \} ~, 
\nonumber \\ 
J^{2-}_{i k_{x}}(\omega_{yz} t) = \left(-\frac{k_{R} \phantom{.} t}{4}\right) 
\{ J_{(i k_{x})-1}(\omega_{yz} t) + J_{(i k_{x})+1}(\omega_{yz} t) 
+ J_{(-i k_{x})-1}(\omega_{yz} t) + J_{(-i k_{x})+1}(\omega_{yz} t) \} ~. 
\label{Kas1002ndOrdRealSolns4Parts}
\end{eqnarray}

Now consider the Bessel time derivative expressions:
\begin{eqnarray}
J_{d+} \equiv \frac{t}{k_{x}} [J_{i k_{x}}(\omega_{yz} t)_{, \phantom{.} t}] 
= \left(\frac{k_{R} \phantom{.} t}{2}\right) 
\{ J_{(i k_{x})-1}(\omega_{yz} t) - J_{(i k_{x})+1}(\omega_{yz} t) \} ~, 
\nonumber \\ 
J_{d-} \equiv \frac{t}{k_{x}} [J_{-i k_{x}}(\omega_{yz} t)_{, \phantom{.} t}] 
= \left(\frac{k_{R} \phantom{.} t}{2}\right) 
\{ J_{(-i k_{x})-1}(\omega_{yz} t) - J_{(-i k_{x})+1}(\omega_{yz} t) \} ~. 
\label{Kas1004thOrdDerivs4Parts}
\end{eqnarray}
These two expressions are clearly complex conjugates of one another, 
thus we can take the real combinations: 
$J^{4+}_{i k_{x}}(\omega_{yz} t) \equiv [(J_{d+} + J_{d-})/2]$, 
$J^{4-}_{i k_{x}}(\omega_{yz} t) \equiv [(J_{d+} - J_{d-})/(2i)]$; 
or, written out in full:
\begin{eqnarray}
J^{4+}_{i k_{x}}(\omega_{yz} t) = \left(\frac{k_{R} \phantom{.} t}{4}\right) 
\{ J_{(i k_{x})-1}(\omega_{yz} t) - J_{(i k_{x})+1}(\omega_{yz} t) 
+ J_{(-i k_{x})-1}(\omega_{yz} t) - J_{(-i k_{x})+1}(\omega_{yz} t) \} ~, 
\nonumber \\ 
J^{4-}_{i k_{x}}(\omega_{yz} t) = \left(\frac{k_{R} \phantom{.} t}{4i}\right) 
\{ J_{(i k_{x})-1}(\omega_{yz} t) - J_{(i k_{x})+1}(\omega_{yz} t) 
- J_{(-i k_{x})-1}(\omega_{yz} t) + J_{(-i k_{x})+1}(\omega_{yz} t) \} ~. 
\label{Kas1004thOrdRealSolns4Parts}
\end{eqnarray}
We thus take $(J^{2+}, J^{2-}, J^{4+}, J^{4-})$ as our basis set 
of real solutions for the Kasner $(1,0,0)$ case.

Now, it is not enough to specify the form of these temporal solutions, 
but we must know the amplitudes of all terms as well (for general 
propagation wavevector $\bm{k}$), to ensure that we have self-consistent 
solutions to be used to form propagating waves. We can obtain 
these amplitudes using the constraints from applying the 
1\textsuperscript{st}-order Maxwell equations -- 
i.e., $\nabla \cdot \bm{E}^{'} = \nabla \cdot \bm{B}^{'} = 0$, 
and Eq's.~\ref{CurlReDefs} (with cyclic permutations), as applied for 
$(p_{x},p_{y},p_{z}) = (1,0,0)$. 

To form waves, we must use combinations of the temporal solutions 
with the spatial functions (recall Eq.~\ref{SepVars}), $S_{i} (x,y,z)$. 
By analogy, in flat spacetime one defines forward and backward propagating 
waves via $\cos [k t \mp (k_{x} x + k_{y} y + k_{z} z)] \equiv 
\cos [k t \mp (\bm{k} \cdot \bm{r})] = 
[\cos(k t) \cos(\bm{k} \cdot \bm{r}) \pm \sin(k t) \sin(\bm{k} \cdot \bm{r})]$, 
with $k = (k_{x}^{2} + k_{y}^{2} + k_{z}^{2})^{1/2}$. 
We can do the same thing here with combinations like 
$[J^{2+}_{i k_{x}}(\omega_{yz} t) \cos(\bm{k} \cdot \bm{r}) 
\pm J^{2-}_{i k_{x}}(\omega_{yz} t) \sin(\bm{k} \cdot \bm{r})]$, and so on 
(still using expressions like $(\bm{k} \cdot \bm{r})$ to refer to the 
flat spacetime version of the dot product). Numerical calculations performed 
by this author show that these are not precisely the exact 
forward/backward-propagating combinations, but have some standing wave 
admixture -- something that can be numerically removed using the late-time, 
almost purely sinusoidal behavior of the Bessel solutions. But for our 
purposes here (defining real analytical solutions) these are adequate; 
we just make sure to note that they produce slightly mixed linear 
combinations of the pure forward/backward waves when used in this way.

Next, we recover the observable electromagnetic fields by undoing 
the transformation from Eq.~\ref{PrimeFieldsReDef}, to transform 
$\bm{E}^{'} \rightarrow \bm{E}$, and $\bm{B}^{'} \rightarrow \bm{B}$. 
We assume the wavenumbers $(k_{x},k_{y},k_{z})$ to be real and nonzero 
(though not necessarily positive). After substantial work to complete 
the aforementioned steps -- including an explicit verification that 
all of the Maxwell and wave equations described above (as applied to 
this Kasner case) are satisfied -- we obtain the (real by construction) 
general solutions as:
\begin{subequations}
\label{Kas100WaveSolns}
\\
\\
Polarization $X_{E}$, forward/backward propagation: 
\begin{eqnarray}
\{E_{x}, E_{y}, E_{z}\} & = & E^{\textrm{f/b}}_{0} ~ 
[J^{2+}_{i k_{x}}(\omega_{yz} t) \cos(\bm{k} \cdot \bm{r} + \phi^{\textrm{f/b}}_{E}) 
\pm J^{2-}_{i k_{x}}(\omega_{yz} t) \sin(\bm{k} \cdot \bm{r}+ \phi^{\textrm{f/b}}_{E})]
\nonumber \\
& \times & \{(t \omega_{yz}^{2}), (-\frac{k_{x} k_{y}}{t}), (-\frac{k_{x} k_{z}}{t})\} ~, 
\nonumber \\ 
\{B_{x}, B_{y}, B_{z}\} & = & E^{\textrm{f/b}}_{0} ~ 
[J^{4+}_{i k_{x}}(\omega_{yz} t) \sin(\bm{k} \cdot \bm{r} + \phi^{\textrm{f/b}}_{E}) 
\mp J^{4-}_{i k_{x}}(\omega_{yz} t) \cos(\bm{k} \cdot \bm{r}+ \phi^{\textrm{f/b}}_{E})]
\nonumber \\
& \times & \{0, (-\frac{k_{x} k_{z}}{t}), (\frac{k_{x} k_{y}}{t})\} ~;
\label{Kas100WaveSolnEfields}
\end{eqnarray}
Polarization $X_{B}$, forward/backward propagation: 
\begin{eqnarray}
\{E_{x}, E_{y}, E_{z}\} & = & B^{\textrm{f/b}}_{0} ~ 
[J^{4+}_{i k_{x}}(\omega_{yz} t) \sin(\bm{k} \cdot \bm{r} + \phi^{\textrm{f/b}}_{B}) 
\mp J^{4-}_{i k_{x}}(\omega_{yz} t) \cos(\bm{k} \cdot \bm{r}+ \phi^{\textrm{f/b}}_{B})]
\nonumber \\
& \times & \{0, (\frac{k_{x} k_{z}}{t}), (-\frac{k_{x} k_{y}}{t})\} ~,
\nonumber \\ 
\{B_{x}, B_{y}, B_{z}\} & = & B^{\textrm{f/b}}_{0} ~ 
[J^{2+}_{i k_{x}}(\omega_{yz} t) \cos(\bm{k} \cdot \bm{r} + \phi^{\textrm{f/b}}_{B}) 
\pm J^{2-}_{i k_{x}}(\omega_{yz} t) \sin(\bm{k} \cdot \bm{r}+ \phi^{\textrm{f/b}}_{B})] 
\nonumber \\
& \times & \{(t \omega_{yz}^{2}), (-\frac{k_{x} k_{y}}{t}), (-\frac{k_{x} k_{z}}{t})\} ~, 
\label{Kas100WaveSolnBfields}
\end{eqnarray}
\end{subequations}
where $\{E^{\textrm{f}}_{0}, E^{\textrm{b}}_{0}, 
B^{\textrm{f}}_{0}, B^{\textrm{b}}_{0}\}$ 
are arbitrary constant amplitudes, and 
$\{\phi^{\textrm{f}}_{E}, \phi^{\textrm{b}}_{E}, 
\phi^{\textrm{f}}_{B}, \phi^{\textrm{b}}_{B}\}$ 
are arbitrary constant phases, as determined by the initial conditions.

(Note that if any of the wavevector components, $(k_{x},k_{y},k_{z})$, 
happen to be zero -- such as for propagation along an axis, or within 
a plane -- then this simplifies matters, since we could use either 
Eq.~\ref{ExWaveEqnWay2} or \ref{ExWaveEqnWay3} to make the right-hand side 
of the 2\textsuperscript{nd}-order wave equation for some of the 
fields go to zero, making the number of solutions for each drop from $4$ to $2$. 
For example, $k_{z} = 0$ means that $B^{'}_{x,\phantom{.} z} = 0$ 
(and $E^{'}_{x,\phantom{.} z} = 0$), so that cyclic permutation of 
Eq.~\ref{ExWaveEqnWay2} leads to a wave equation for $E^{'}_{y,\phantom{.} z}$ 
(and $B^{'}_{y,\phantom{.} z}$) with no driving term and only $2$ solutions, 
$J^{2\pm}$. And this works similarly with other restricted propagation directions, 
leading to appropriately simplified versions of the most general solutions 
given above in Eq's.~\ref{Kas100WaveSolns}.)

To study the energetics of these solutions, we compute the electromagnetic 
stress-energy tensor via \citep[e.g.,][]{WeinGC}: 
$T^{\mu \nu} = F^{\mu}_{\phantom{0.} \lambda} F^{\nu \lambda} 
- (g^{\mu \nu} F_{\alpha \beta} F^{\alpha \beta})/4$. 
Following Misner {\it et al.} \cite{MTW}, we write the four-momentum density 
per unit volume (as measured in an observer's local Lorentz frame) as: 
$d P^{\mu}/dV = -T^{\mu \nu} g_{\nu \gamma} U^{\gamma}$; 
and by choosing a stationary, comoving observer in the Kasner metric, 
we have $U^{\gamma} = (1,0,0,0)$. Then, following Weinberg \cite{WeinGC}, 
we write the total integrated four-momentum with a specified volume 
as $P^{\mu} = \int (d P^{\mu}/dV) 
\sqrt{-|g_{\mu \nu}|} d^{\phantom{.} 3}x 
= [t (d P^{\mu}/dV) \Delta x \Delta y \Delta z] $.

For simplicity, we restrict the calculation to a single component: 
the forward-propagating $X_{E}$ polarization solution. Even so, 
the results are too complicated to easily evaluate analytically. 
Instead we study the behavior numerically, selecting a variety of 
(order-unity) values for $(k_{x}, k_{y}, k_{z})$, adopting some 
particular value of $(x,y,z)$ in this (homogeneous) model, and examining 
the behaviors of the energy and momenta $P^{\mu}$ over time.

At large $t$, these $P^{\mu}$ all oscillate with a period 
of $\sim (\pi/\sqrt{k_{y}^{2}+k_{z}^{2}})$, which makes sense because 
equations like Eq.~\ref{Kas1002ndOrdExBx} become 
$F^{''}_{x,\phantom{.} t,\phantom{.} t} 
+ (k_{y}^{2} + k_{z}^{2}) F^{''}_{x} \approx 0$, 
with obvious solutions $\sim \cos/\sin[(k_{y}^{2} + k_{z}^{2})^{1/2} t]$.

As small $t$, however, the oscillation frequency increases at 
an increasing rate and without bound, leading to a self-similar behavior 
with an infinite number of oscillations as $t \rightarrow 0$. 
Analytically this makes sense because the $z \rightarrow 0$ limit of 
a Bessel function (at fixed order) is known \cite{AbramStegun} to be 
$J_{\nu}(z) \approx (z/2)^{\nu}/\Gamma (\nu + 1)$, and thus for 
very small $t$ we can approximate:
\begin{eqnarray}
J_{(\pm i k_{x})}(\omega_{yz} t) & \approx & 
\frac{(\omega_{yz} t /2)^{\pm i k_{x}}}{\Gamma (\pm i k_{x} + 1)}
\propto (\omega_{yz} t /2)^{\pm i k_{x}} \nonumber \\ 
& = & \exp{\{\ln{[(\omega_{yz} t /2)^{\pm i k_{x}}]}\}} 
= \exp{[(\pm i k_{x}) \ln{(\omega_{yz} t /2)}]} \nonumber \\ 
& = & \cos{[k_{x} \ln{(\omega_{yz} t /2)}]} 
\pm \sin{[k_{x} \ln{(\omega_{yz} t /2)}]} ~, 
\label{SmalltKas100}
\end{eqnarray}
justifying the self-similar, infinitely oscillatory behavior described 
above for $t \rightarrow 0$. This result is consistent 
with the analysis in Sagnotti and Zwiebach \cite{SagZwiebachWaves}, 
where they found an infinite number of phase oscillations to occur as 
$t \rightarrow 0$; and in our numerical studies, we similarly 
find oscillations (varying $\sim 20-30\%$ for order-unity $k_{i}$ choices) 
in the phase velocity of the fields satisfying Eq.~\ref{Kas1002ndOrdExBx} 
as $t \rightarrow 0$. Zooming in on the small-$t$ regime, those oscillations 
also seem to be self-similar in character, becoming increasingly frequent 
as we get to the smallest simulated values of $t$.

Now considering the overall amplitude envelope of these oscillating 
$P^{\mu}$, we find that $T^{0y} \sim T^{0z} \propto t^{-1}$, so that 
$P^{y} \sim P^{z}$ are constant in time, which makes physical sense 
given that the $x$- and $y-$axes are static, and it agrees with 
Equation 3.9 of Sagnotti and Zwiebach \cite{SagZwiebachWaves}. 
However, we find that $T^{0x} \propto t^{-3}$ (such that $P^{x} \propto t^{-2}$), 
which is not predicted by their Equation 3.9, and reflects the 
evolving nature of the $x$-axis, the only non-static direction.
A result reminiscent of this can, in fact, be seen in their 
Equation 3.10, where their $T^{00}$ energy expression contains 
terms proportional both to $t^{-1}$ and to $t^{-3}$; though that 
expression was obtained for $t \rightarrow \infty$ (where the 
WKB/geometric optics approximation holds), and ignoring interference terms. 
From our explicit calculation, at large $t$ -- 
where $P^{y}$ and $P^{z}$ dominate -- we find that $T^{00} \propto t^{-1}$, 
just like $T^{0y}$ and $T^{0z}$, so that the envelope of $P^{t}$ is constant; 
but for $t \rightarrow 0$, where $P^{x}$ dominates, $T^{00}$ makes 
a sharp turn at some small, critical value of $t$ (determined by the 
$k_{i}$ parameters), and its envelope becomes proportional to $t^{-2}$ 
for smaller $t$ values, so that $P^{t} \propto t^{-1}$. 
(These results also seem generally in line with those of 
Goorjian \cite{GoorjianKas}; though they restricted their 
electromagnetic vector potential fields to lie along a specific 
spatial axis, a condition which we have not imposed here.)

Lastly for this case, considering that $T^{00} \propto t^{-2}$ at 
small $t$, there may be concern that the energy density increases 
so much that the ``test field" approximation mentioned in 
Section~\ref{KasMet} might break down back towards the initial singularity. 
However, for Kasner cases with general indices $(p_{x},p_{y},p_{z})$, 
one finds that the Ricci tensor components (and the Ricci scalar) due to 
the vacuum gravitational fields are also proportional to $t^{-2}$, 
so that they increase just as fast as this electromagnetic $T^{00}$ does 
as $t \rightarrow 0$, thus the gravitational feedback by the 
electromagnetic fields would remain equally unimportant (relative to that 
of the vacuum gravitational fields) at all times. But before one concludes 
that the test field assumption should never be a problem for stress-energy terms 
with this time dependence, one must recall that this particular 
Kasner $(1,0,0)$ case is actually flat spacetime, and thus the Ricci 
tensor and scalar are always zero, apparently making this case somewhat 
ambiguous. (In fact, in Appendix~\ref{Kas100Mink} below, we show that this 
problem transforms exactly to the case of the ordinary fields oscillating 
on a Minkowski background.) However, this also allows us to draw the obvious 
conclusion that as long as one does not {\it start out} with electromagnetic 
waves of such an intensity that they would create significant curvature even 
in flat space, then the test field assumption should remain a safe one here.

\section{Kasner special case $\bm{\{(-1/3),(2/3),(2/3)\}}$}
\label{SecKas2323N13}
The other axisymmetric case satisfying the vacuum Kasner conditions 
is the one where two axes expand as $\sim t^{2/3}$, with the third
contracting as $\sim t^{-1/3}$. Sticking to the convention of choosing 
the $x$-axis as the one with the ``unique" expansion rate, 
we define $\{p_{x}, p_{y}, p_{z}\} = \{(-1/3),(2/3),(2/3)\}$. 

Applying these indices to Eq.~\ref{GenKas2ndOrdkxkykz}, the 
2\textsuperscript{nd}-order wave equation 
for $F^{''}_{x} \in \{E^{''}_{x} (t)$, $B^{''}_{x} (t)\}$ becomes:
\begin{equation}
F^{''}_{x,\phantom{.} t,\phantom{.} t} 
- \left( \frac{1}{3 \phantom{.} t} \right) F^{''}_{x,\phantom{.} t} 
+ \left( k_{x}^{2} \phantom{.} t^{2/3} + \frac{k_{y}^{2} + k_{z}^{2}}{t^{4/3}} 
\right) F^{''}_{x} = 0 ~, \label{Kas2323n132ndOrdExBx}
\end{equation}

Alternatively, the temporal functions for the remaining fields, 
$F^{''}_{y} \in \{E^{''}_{y} (t), B^{''}_{y} (t)\}$ 
and $F^{''}_{z} \in \{E^{''}_{z} (t), B^{''}_{z} (t)\}$, 
will satisfy 4\textsuperscript{th}-order wave equations derived 
from appropriate cyclic permutations of Eq.~\ref{GenKas4thOrderDiffEq} 
for these $p_{i}$ values.

As we saw for the Kasner $(1,0,0)$ special case, the easiest way to solve 
the 4\textsuperscript{th}-order equations would be by guessing 
modified versions of the solutions to the homogeneous 
2\textsuperscript{nd}-order equation. Unfortunately for this case, 
Eq.~\ref{Kas2323n132ndOrdExBx} does not appear to be a 
simply transformable variation of any standard differential equation 
that has known analytical solutions. (In general, it does not appear 
that there are known analytical solutions to this type of equation 
when one has at least two distinct $p_{i}$ indices that are each 
unequal to $1$.)

It is known \citep[e.g.,][]{OlPetKasner}, as can easily be determined 
using standard mathematical software, that solutions to equations 
like Eq.~\ref{Kas2323n132ndOrdExBx} can be expressed as non-resolved 
integrals of Heun Biconfluent functions; but it is unclear if such 
expressions are more useful than just obtaining direct numerical solutions 
to the equation. In any case, this author has not found any convenient 
analytical solutions to the 2\textsuperscript{nd}-order wave equation 
for this Kasner case.

This problem has long been a stumbling block for researchers studying this 
(and more complicated) Kasner cases, so that various simplifications or 
approximations are necessary to obtain some kind of analytical understanding 
and solutions. As discussed above in the Introduction, a number of authors 
have opted either to restrict the wave propagation to be along a single spatial 
axis \cite{AsenjoHoj,GoorjianKas,SagZwiebachWaves}, to focus on early- and 
late-time approximations \cite{DhurVishVeshCohen, PonsMarciSeries, AnsaryFieldSeries}, 
or both \cite{MatWollKasEM}. Previously, we have obtained solutions for 
specific cases involving restricted propagation wavevectors in metrics with 
carefully chosen axis expansion rates \cite{BochProc28Tex}; 
but as we are mainly interested in the most general propagation behaviors, 
we will primarily employ fully numerical methods for these studies going forward.

One interesting use of numerical methods here is to study the phase velocity 
of the propagating $F^{''}_{x}$ fields obeying Eq.~\ref{Kas2323n132ndOrdExBx}. 
Some of the results can be predicted from the easily obtained early- and 
late-time solutions, as are given (for example) in 
Dhurandhar {\it et al.} \cite{DhurVishVeshCohen}; while some other results
are discovered numerically. But considering the approximated solutions 
first can provide insight for interpreting the full numerical solutions. 

For late times, where $[(k_{y}^{2} + k_{z}^{2}) \phantom{.} t^{-4/3}] << 
(k_{x}^{2} \phantom{.} t^{2/3})$, we drop the former term from 
Eq.~\ref{Kas2323n132ndOrdExBx} -- essentially equivalent to setting 
$k_{y}^{2} \approx k_{z}^{2} \approx 0$ -- and the ``exact" solution 
for that approximated equation is a linear combination of sinusoids, 
$F^{''}_{x} \propto \cos / \sin [(3/4) \phantom{.} k_{x} \phantom{.} t^{4/3}]$. 
The implies an adiabatically-varying temporal oscillation frequency of 
$\omega_{\textrm{a}} 
\equiv [(3/4) \phantom{.} k_{x} \phantom{.} t^{4/3}]_{,\phantom{.} t} 
= (k_{x} \phantom{.} t^{1/3})$, which makes sense given that 
Eq.~\ref{Kas2323n132ndOrdExBx} becomes 
$F^{''}_{x,\phantom{.} t,\phantom{.} t} \approx 
- [(k_{x} \phantom{.} t^{1/3})^{2}] F^{''}_{x}$ for very large $t$.
Since the solution is sinusoidal, we can read off the 
(coordinate) phase velocity of this late-time wave as 
$v \approx v_{x} \approx \omega_{\textrm{a}} / k \approx \omega_{\textrm{a}} / k_{x} 
= (k_{x} \phantom{.} t^{1/3}) / k_{x} = t^{1/3}$ (and where we recall 
that $c \equiv 1$). Thus the late-time physical speed (aligned almost 
entirely along the contracting x-axis) is equal to 
$v^{\textrm{Phys}} \approx g_{xx} (v_{x})^{2} = t^{-2/3} (t^{1/3})^{2} = 1$. 
Recalling also that $g_{tt} = -1$, the wave thus behaves now exactly as a 
light ray would, propagating at null speed and obeying the geometric optics 
approximation, just as one expects to be true when the frequencies 
and/or time $t$ are sufficiently large \cite{DhurVishVeshCohen}. 

%(Note that we set $k \rightarrow k_{x}$ in the above argument, 
%since nonzero $(k_{y}, k_{z})$ is mathematically inconsistent with the 
%``exact" sinusoidal solution quoted above; but numerical studies of the 
%full equation verify that $(v_{y}, v_{z})$ do in fact diminish at large $t$, 
%and $v \rightarrow v_{x} \rightarrow 1$, even for nonzero $(k_{y}, k_{z})$.)

We cannot, however, assume geometric optics behavior all the way down 
to $t \rightarrow 0$, where the terms containing negative powers of $t$ 
and the first time derivative of the field are important. For early times, where 
$[(k_{y}^{2} + k_{z}^{2}) \phantom{.} t^{-4/3}] >> (k_{x}^{2} \phantom{.} t^{2/3})$, 
we drop the latter term, and the ``exact" solution to 
Eq.~\ref{Kas2323n132ndOrdExBx} approximated in this way involves linear 
combinations of Bessel and Neumann functions of order $2$, becoming: 
$F^{''}_{x} \propto \{ (t^{2/3}) \phantom{.} J_{2} / Y_{2} 
[3 \phantom{.} \sqrt{(k_{y}^{2} + k_{z}^{2})} \phantom{.} t^{1/3}] \}$. 

The argument of these solutions looks right, giving an effective 
(when adiabatically-varying) frequency of $\omega_{\textrm{a}} 
\equiv [3 \phantom{.} \sqrt{(k_{y}^{2} + k_{z}^{2})} 
\phantom{.} t^{1/3}]_{,\phantom{.} t} 
= [\sqrt{(k_{y}^{2} + k_{z}^{2})} \phantom{.} t^{-2/3}]$, 
which makes sense to the extent that Eq.~\ref{Kas2323n132ndOrdExBx} 
for early times may {\it almost} be approximated as 
$F^{''}_{x,\phantom{.} t,\phantom{.} t} \approx 
- \{[\sqrt{(k_{y}^{2} + k_{z}^{2})} 
\phantom{.} t^{-2/3}]^{ 2}\} F^{''}_{x}$.
But in this case, the $[F^{''}_{x,\phantom{.} t}/(3t)]$ term 
remains important, and the Bessel/Neumann function solutions here will 
deviate significantly from sinusoidal propagating behavior at 
small-$t$. (Dropping the $[F^{''}_{x,\phantom{.} t}/(3t)]$ term as well, 
of course, would finally result in sinusoidal solutions with argument 
proportional to $t^{1/3}$; but numerical work confirms that the Bessel/Neumann 
functions are actually the correct (approximate) solutions down to 
$t \rightarrow 0$, as will be demonstrated shortly.) Hence we expect 
(and indeed find) the wave phase velocity to differ significantly 
from the geometric optics expectation for null rays as $t \rightarrow 0$; 
namely, failing to evolve as $\{v_{y}, v_{z}\} \propto t^{-2/3}$. 
(Though $v_{x} \approx 0$ still remains true).

In Section~\ref{SecKas100} above, we briefly discussed the construction 
of purely forward- and backward-propagating waves from the known 
analytical temporal solutions for the Kasner $(1,0,0)$ case, in conjunction 
with the sinusoidal spatial functions. Using the almost purely sinusoidal 
behavior of Bessel functions at large-$t$, we combined the 
spatial and temporal functions in a way that numerically eliminated 
nearly all of the standing wave contributions for the forward (or backward) 
-- i.e., rightward (or leftward) -- traveling waves. From those constructed 
unidirectional waves, we obtained the phase velocity of each by following a 
wavefront of its $\{E^{'}_{x} (t,x,y,z)$, $B^{'}_{x} (t,x,y,z)\}$ fields, 
after which we used the metric to calculate the true physical speed of that 
wave phase velocity. As noted in that section, our simulations for the 
Kasner $(1,0,0)$ case found phase velocities that oscillated 
(about the null speed of $v^{\textrm{Phys}} = 1$), in a presumably infinite 
series of self-similar oscillations as $t \rightarrow 0$; results that were 
in qualitative agreement with findings from prior authors.

For this Kasner $\{(-1/3),(2/3),(2/3)\}$ case, we conduct a similar procedure; 
but, not having analytical solutions for the $F^{''}_{x}$ fields obeying the full 
Eq.~\ref{Kas2323n132ndOrdExBx}, we used a pair of numerical solutions (similarly 
treated to remove standing-wave contributions) in order to construct the rightward- 
and leftward-propagating waves. (Note that the waves propagating in either 
direction behaved indistinguishably from one another, as they should for this 
spatially homogeneous spacetime.) But what we do find for this metric, is that 
wave propagation speeds are {\it inhibited} as the initial singularity is approached. 
The phase velocity of the waves, though staying very close to the null ray speed of 
$v^{\textrm{Phys}} = 1$ for sufficiently large $t$, begins to decrease for $t$ values 
close to unity (the transition time depending quantitatively upon the particular 
$k_{i}$ values), and as $t \rightarrow 0$ here, $v^{\textrm{Phys}} \rightarrow 0$. 

\begin{figure}
\includegraphics{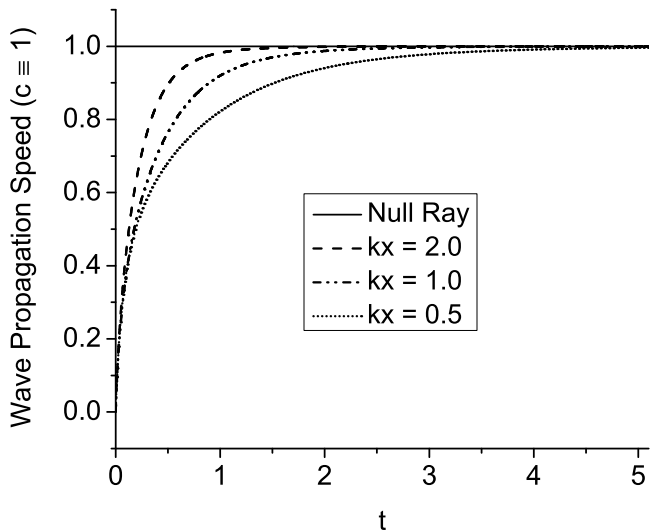}
\caption{Total physical speed of the phase velocity 
for the $\{E^{'}_{x}$, $B^{'}_{x}\}$ fields obeying 
Eq.~\ref{Kas2323n132ndOrdExBx} in the Kasner $\{(-1/3),(2/3),(2/3)\}$ 
metric, plotted versus time from the initial singularity. 
The wavenumber $k_{x}$ is varied, while holding $(k_{y}^{2} + k_{z}^{2}) = 1$. 
For each numerical simulation, rightward and leftward propagating waves for 
each case are found to produce indistinguishable results. The line $v = 1$ 
represents the prediction for null rays according to geometric optics.
\label{WaveSpdPlotK2323n13}}
\end{figure}

A plot of this (physical) wave phase velocity is shown in 
Figure~\ref{WaveSpdPlotK2323n13}, for three sets of wavevector values: 
$k_{x} = \{0.5, 1.0, 2.0 \}$, where for all of those we set 
$k_{yz} \equiv \sqrt{k_{y}^{2} + k_{z}^{2}} = 1$. As expected, 
we see that higher frequencies (i.e., larger $k_{x}$ here) insures 
better agreement (down to earlier time $t$) with the $v = c$ light ray 
speed expectation; but in all cases, as $t \rightarrow 0$, there is a 
point where the predictions of geometric optics break down, and the 
phase velocity of the waves (in all propagation directions, as our 
numerical results show) become inhibited by the anisotropically-contracting 
nature of this metric. While it is difficult to make conclusions about 
energy propagation speeds based solely upon phase velocities, a 
naive conclusion would be that energy propagation may very well 
get choked off as $t \rightarrow 0$ in this metric; a purely 
wave-based effect that has no analogous implication from 
simple null-ray estimations.

Beyond the obvious interest in such findings obtained for these homogeneous 
vacuum metrics, it is especially intriguing that the 
Kasner $\{(-1/3),(2/3),(2/3)\}$ case is equivalent to the vacuum-limit 
of a radiation-filled metric that is not just anisotropic, but also 
inhomogeneous; yet which remains highly symmetrical nevertheless, 
and thus amenable for study.

Specifically, what may be called the Kuang-Li-Liang (KLL) metric -- 
i.e., one particular variety of the cases given in 
Kuang {\it et al.} \cite{KLLmetricPaper} -- is defined as: 
\begin{equation}
ds^{2} = \frac{J^{2}(T \pm X)}{T^{1/2}} (-dT^{2} + dX^{2}) + T (dY^{2} + dZ^{2}) ~, 
\label{KLLmetricDefn}
\end{equation}
where $J$ is a general (real) function of the variable $(T+X)$ -- 
or alternatively, of $(T-X)$ -- which to satisfy reasonable energy conditions 
must obey $J^{'}/J > 0$. This metric breaks homogeneity but remains 
planar-symmetric gravitationally, and is filled with 
pure electromagnetic radiation. (That background radiation is 
semi-plane-symmetric, in the sense that rotations within the $YZ$-plane 
will alter the electromagnetic polarization but not its energy density.) 

This non-vacuum, conformally non-flat metric also possesses a cosmological 
(expanding/contracting) quality, in that it is known \citep{LiLiangToKasner} 
that the limit $J \rightarrow 1$ makes this metric exactly equivalent to the 
Kasner $\{(-1/3),(2/3),(2/3)\}$ case. One can verify this using the 
substitutions: $t = [(4/3) \phantom{.} T^{3/4}]$, $x = [(3/4)^{-1/3} \phantom{.} X]$, 
$y = [(3/4)^{2/3} \phantom{.} Y]$, $z = [(3/4)^{2/3} \phantom{.} Z]$.

Such properties all make this KLL metric (for various choices of $J(T \pm X)$) 
an extremely interesting physical system for studying wave propagation, using 
the formalism, methodology, and numerical tools developed for and demonstrated 
in this paper. As a generalization of Kasner $\{(-1/3),(2/3),(2/3)\}$, 
the KLL metric almost certainly needs a numerical treatment; and we intend to 
use the intuition gained from all of our results from the Kasner cases 
discussed here as a stepping stone for interpreting future results that will be 
obtained for that physically richer (though more mathematically complicated) system.

\section{Discussion and summary}
\label{SecConclude}
In this paper, we applied the curved spacetime Maxwell equations 
to the general class of Kasner metrics, and used the usual 
vector identities to produce 2\textsuperscript{nd}-order 
differential equations for the full set of electromagnetic fields.

Unlike the results from previous authors that we have seen, 
our 2\textsuperscript{nd}-order wave equations were not fully uncoupled 
(as they are in flat spacetime), but contained nonhomogeneous driving terms 
indicating a mixing between the electric and magnetic field components, 
generated directly by the anisotropic nature of the Kasner 
expansion/contraction rates. To eliminate the coupling in the wave equations, 
it was necessary to produce 4\textsuperscript{th}-order differential equations; 
and consequently we derived 4\textsuperscript{th}-order wave equations 
that are valid for all of the fields in any metric with Kasner-like 
axis expansion rate coefficients (i.e., whether or not they satisfy 
the Kasner vacuum conditions). 

We then considered two special axisymmetric Kasner cases, for which the 
wave equations for the fields along the axis of symmetry are greatly simplified. 
First, for the $(p_{x}, p_{y}, p_{z}) = (1,0,0)$ case, we obtained the explicit 
solutions for all of the fields, for a wave with the most general wavevector 
components, traveling in a general direction through the three dimensional space.
This included full solutions to the 4\textsuperscript{th}-order wave equations, 
deriving what we believe are additional field solutions that had not been found 
in previous studies of this Kasner case. We also studied the energetics of the 
fields, showing them to agree (where comparable) with previous studies, and in 
general providing us with confidence in the appropriateness of the assumption 
made by treating these solutions as test fields, propagating upon an 
essentially unchanged background Kasner metric.

Next, we considered the $\{p_{x}, p_{y}, p_{z}\} = \{(-1/3),(2/3),(2/3)\}$ 
case, deriving our 2\textsuperscript{nd}-order wave equations for the fields, 
and pointing out (as noted by previous researchers) that the equations 
(even without considering the nonhomogeneous driving terms) are analytically 
unsolvable. Using late-time and early-time approximations, we set expectations 
for what the temporal oscillation frequencies would be; and we also noted that 
the geometric optics approximation should be valid at late times, but should 
be expected to break down at early times, leading to deviations from null ray 
behavior that only become apparent as $t$ decreases below (approximately) unity 
and heads to zero. 

Using a numerical program written specifically for this research, designed 
to compute phase velocity of the wave by following a wavefront through the evolving 
metric, we find that the above expectations were correct: at late times, the 
physical speed of the phase velocity is nearly exactly equal to $v = c \equiv 1$; 
but as $t \rightarrow 0$, the wave propagation is sharply inhibited, causing 
$v \rightarrow 0$. Again as expected, larger wavenumbers (i.e., larger $k_{x}$) 
allowed the geometric optics approximation to remain valid longer, 
down to smaller $t$, forestalling (but not preventing) the ultimate breakdown 
of the validity of the null ray treatment of the light waves.

Lastly, we noted a very interesting class of non-vacuum, radiation-filled, 
inhomogeneous metrics with cosmology-like behavior, for which the Kasner 
$\{(-1/3),(2/3),(2/3)\}$ case is the vacuum limit; and which should be 
perfectly suited for future study using the analytical formalism and 
numerical tools developed in this paper.

\appendix

\section{Derivation of the 2\textsuperscript{nd}-order wave equation}
\label{App2ndOrderDeriv}
Here we derive the inhomogeneous wave equation for $E^{'}_{x}$, Eq.~\ref{GenKas2ndOrdEx} 
-- from which similar equations can be inferred for all of the fields --  
from the 1\textsuperscript{st}-order curved space Maxwell equations for such fields.

Recall that these renormalized fields obey 
$\nabla \cdot \bm{E}^{'} = \nabla \cdot \bm{B}^{'} = 0$ 
(the dot products being defined here as in flat spacetime); 
and the three curl equations which we need here 
(of the six inferred from Eq's.~\ref{CurlReDefs}), 
are reproduced here as:
\begin{subequations}
\label{CurlEqnsFor2ndOrdEqn}
\begin{equation}
E^{'}_{x,\phantom{.} t} = 
[t^{(-p_{x} - p_{y} + p_{z})} B^{'}_{z,\phantom{.} y}] 
- [t^{(-p_{x} + p_{y} - p_{z})} B^{'}_{y,\phantom{.} z}] 
~, \label{CurlEqnEx}
\end{equation}
\begin{equation}
B^{'}_{y,\phantom{.} t} = 
[t^{(-p_{x} - p_{y} + p_{z})} E^{'}_{z,\phantom{.} x}] -
[t^{(p_{x} - p_{y} - p_{z})} E^{'}_{x,\phantom{.} z}] ~, 
\label{CurlEqnBy}
\end{equation}
\begin{equation}
B^{'}_{z,\phantom{.} t} = 
[t^{(p_{x} - p_{y} - p_{z})} E^{'}_{x,\phantom{.} y}] -
[t^{(-p_{x} + p_{y} - p_{z})} E^{'}_{y,\phantom{.} x}] ~. 
\label{CurlEqnBz}
\end{equation}
\end{subequations}

Next, we take the combination: 
\begin{equation}
\partial/\partial y \{ [t^{(-p_{x} - p_{y} + p_{z})}] \times \textrm{Eq.~\ref{CurlEqnBz}} \} 
- \partial/\partial z \{ [t^{(-p_{x} + p_{y} - p_{z})}] \times \textrm{Eq.~\ref{CurlEqnBy}} \} 
~. \label{CurlItStep1}
\end{equation}
For clarity, we will treat the left hand side (LHS) and right hand side (RHS) 
of the resulting equation separately. Working out the combination just mentioned, 
we get:
\begin{subequations}
\label{CurlItStep2}
\begin{equation}
\textrm{LHS} = 
[t^{(-p_{x} - p_{y} + p_{z})}] B^{'}_{z,\phantom{.} t,\phantom{.} y}
- [t^{(-p_{x} + p_{y} - p_{z})}] B^{'}_{y,\phantom{.} t,\phantom{.} z}
~, \label{CurlItStep2a}
\end{equation}
\begin{equation}
\textrm{RHS} = 
[t^{-2p_{y}} E^{'}_{x}]_{,\phantom{.} y,\phantom{.} y} 
- [t^{-2p_{x}} E^{'}_{y}]_{,\phantom{.} y,\phantom{.} x} 
- [t^{-2p_{x}} E^{'}_{z}]_{,\phantom{.} z,\phantom{.} x} 
+ [t^{-2p_{z}} E^{'}_{x}]_{,\phantom{.} z,\phantom{.} z} 
~. \label{CurlItStep2b}
\end{equation}
\end{subequations}
Note that we have used the fact that we can exchange (commute) 
the spatial partial derivatives freely here.

Then, using $\nabla \cdot \bm{E}^{'} = 0$, and thus 
$[t^{-2p_{x}} (\nabla \cdot \bm{E}^{'})_{,\phantom{.} x}] = 0$, 
we can write:
\begin{equation}
[t^{-2p_{x}} E^{'}_{x}]_{,\phantom{.} x,\phantom{.} x} =
- [t^{-2p_{x}} E^{'}_{y}]_{,\phantom{.} y,\phantom{.} x} 
- [t^{-2p_{x}} E^{'}_{z}]_{,\phantom{.} z,\phantom{.} x} 
~, \label{DivRemove}
\end{equation}
and hence we get:
\begin{equation}
\textrm{RHS} = 
[t^{-2p_{x}} E^{'}_{x}]_{,\phantom{.} x,\phantom{.} x}
+ [t^{-2p_{y}} E^{'}_{x}]_{,\phantom{.} y,\phantom{.} y} 
+ [t^{-2p_{z}} E^{'}_{x}]_{,\phantom{.} z,\phantom{.} z} 
\equiv \{t \nabla ^{2}\} E^{'}_{x} ~. \label{RHSdone}
\end{equation}

Next, for the LHS, we commute the time derivatives outward in 
Eq.~\ref{CurlItStep2a}; though since $\partial/\partial t$ 
cannot be moved through powers of $t$ without generating 
extra product rule terms, we add them in as necessary, to get: 
\begin{eqnarray}
\textrm{LHS} & = & 
\{ [t^{(-p_{x} - p_{y} + p_{z})}] B^{'}_{z,\phantom{.} y}
- [t^{(-p_{x} + p_{y} - p_{z})}] B^{'}_{y,\phantom{.} z} \}_{,\phantom{.} t} 
\nonumber \\ 
& - & \left[\frac{(-p_{x} - p_{y} + p_{z})}{t} t^{(-p_{x} 
- p_{y} + p_{z})} B^{'}_{z,\phantom{.} y}\right] 
+ \left[\frac{(-p_{x} + p_{y} - p_{z})}{t} t^{(-p_{x} 
+ p_{y} - p_{z})} B^{'}_{y,\phantom{.} z}\right] 
\nonumber \\
& = & \{ E^{'}_{x,\phantom{.} t} \}_{,\phantom{.} t} 
+ \left(\frac{p_{x}}{t}\right) \{ [t^{(-p_{x} - p_{y} + p_{z})} B^{'}_{z,\phantom{.} y}] 
- [t^{(-p_{x} + p_{y} - p_{z})} B^{'}_{y,\phantom{.} z}] \}
\nonumber \\
& + & \frac{(p_{y} - p_{z})}{t} [t^{(-p_{x} - p_{y} + p_{z})} B^{'}_{z,\phantom{.} y} 
+ t^{(-p_{x} + p_{y} - p_{z})} B^{'}_{y,\phantom{.} z}]
\nonumber \\
& = & E^{'}_{x,\phantom{.} t,\phantom{.} t}  
+ \left(\frac{p_{x}}{t}\right) E^{'}_{x,\phantom{.} t} 
\nonumber \\
& + & \frac{(p_{y} - p_{z})}{t} [t^{(-p_{x} - p_{y} + p_{z})} B^{'}_{z,\phantom{.} y} 
+ t^{(-p_{x} + p_{y} - p_{z})} B^{'}_{y,\phantom{.} z}]
~, \label{LHSmovingDDt}
\end{eqnarray}
where the second and third equalities were obtained using 
repeated applications of Eq.~\ref{CurlEqnEx}. 

Finally, setting the LHS from Eq.~\ref{LHSmovingDDt} equal to the RHS 
from Eq.~\ref{RHSdone}, we get:
\begin{equation}
E^{'}_{x,\phantom{.} t,\phantom{.} t}
+ \left(\frac{p_{x}}{t}\right) E^{'}_{x,\phantom{.} t} 
+ \frac{(p_{y} - p_{z})}{t} [t^{(-p_{x} - p_{y} + p_{z})} B^{'}_{z,\phantom{.} y} 
+ t^{(-p_{x} + p_{y} - p_{z})} B^{'}_{y,\phantom{.} z}] 
= \{t \nabla ^{2}\} E^{'}_{x} ~, \label{Ex2ndOrdEqnDone}
\end{equation}
which is the same as Equations~\ref{GenKas2ndOrdEx},\ref{ExWaveEqnWay1}, as promised.

\section{Reduction of the Kasner $\bm{(1,0,0)}$ electromagnetic tensor 
to Minkowski Form}
\label{Kas100Mink}
It is known \cite{LandauLifClassFields} that the metric in Eq.~\ref{Kas100Metric} 
can be transformed to Minkowski space with the substitutions: 
\begin{equation}
\tilde{t} \equiv t \cosh{x} ~,~ \tilde{x} = t \sinh{x} ~.
\label{FlatKas100transform}
\end{equation}
The reverse transformations are thus given by:
\begin{equation}
t = (\tilde{t}^{2} - \tilde{x}^{2})^{1/2} ~,~ x = \arctanh{(\tilde{x}/\tilde{t})} ~.
\label{FlatKas100revtransform}
\end{equation}

The covariant electromagnetic tensor can be transformed to the new coordinate system 
via \cite{WeinGC}:
\begin{equation}
\tilde{F}^{\mu \nu} = \frac{\partial \tilde{x}^{\mu}}{\partial x^{\alpha}} 
\frac{\partial \tilde{x}^{\nu}}{\partial x^{\beta}} F^{\alpha \beta} ~,
\label{EmTensorTransformDiffs}
\end{equation}
with:
\begin{equation}
\mathcal{M_{K}} \equiv 
\frac{\partial \tilde{x}^{\mu}}{\partial x^{\alpha}} =  
\begin{bmatrix}
\cosh{x} & t \sinh{x} & 0 & 0 \\
\sinh{x} & t \cosh{x} & 0 & 0 \\
 0 & 0 & 1 & 0 \\
 0 & 0 & 0 & 1 \\
\end{bmatrix} 
= 
\begin{bmatrix}
\tilde{t} (\tilde{t}^{2} - \tilde{x}^{2})^{-1/2} & \tilde{x} & 0 & 0 \\
\tilde{x} (\tilde{t}^{2} - \tilde{x}^{2})^{-1/2} & \tilde{t} & 0 & 0 \\
 0 & 0 & 1 & 0 \\
 0 & 0 & 0 & 1\\
\end{bmatrix} 
~.\label{EmTensorTransformMat}
\end{equation}

As defined in Section~\ref{EMconv}, and applied to this Kasner $(1,0,0)$ 
metric, we can write the (contravariant) electromagnetic tensor as: 
\begin{equation}
F^{\alpha \beta} = 
\begin{bmatrix}
0 & (t^{-2} E_{x}) & E_{y} & E_{z} \\
-(t^{-2} E_{x}) & 0 & (t^{-1} B_{z}) & -(t^{-1} B_{y}) \\
-E_{y} & -(t^{-1} B_{z}) & 0 & (t^{-1} B_{x}) \\
-E_{z} & (t^{-1} B_{y}) & -(t^{-1} B_{x}) & 0 \\
\end{bmatrix}
~.\label{Kas100EmContraFab}
\end{equation}

Now, remembering that the new coordinates are Minkowski space -- 
so that we raise and lower indices with $\eta = Diag(-1,1,1,1)$ -- 
the transformed electromagnetic (covariant) tensor is calculated as:
\begin{eqnarray}
\tilde{F}_{\mu \nu} & = & \eta \tilde{F}^{\mu \nu} (\eta)^{T} 
= \eta \mathcal{M_{K}} F^{\alpha \beta} 
(\eta \mathcal{M_{K}})^{T} \nonumber \\
& = & \frac{1}{(\tilde{t}^{2} - \tilde{x}^{2})^{1/2}}
\begin{bmatrix}
0  & -E_{x} & -(E_{y} \tilde{t} + B_{z} \tilde{x}) & -(E_{z} \tilde{t} - B_{y} \tilde{x}) \\
E_{x} & 0 & (B_{z} \tilde{t} + E_{y} \tilde{x}) & -(B_{y} \tilde{t} - E_{z} \tilde{x}) \\
(E_{y} \tilde{t} + B_{z} \tilde{x}) & -(B_{z} \tilde{t} + E_{y} \tilde{x}) & 0 & B_{x} \\
(E_{z} \tilde{t} - B_{y} \tilde{x}) & (B_{y} \tilde{t} - E_{z} \tilde{x}) & -B_{x} & 0 \\
\end{bmatrix} 
~.\label{MinkEmTensorMixed}
\end{eqnarray}

Comparing this result to the corresponding flat spacetime tensor:
\begin{equation}
\tilde{F}^{\textrm{(Flat)}}_{\mu \nu} = 
\begin{bmatrix}
0 & -\tilde{E}_{x} & -\tilde{E}_{y} & -\tilde{E}_{z} \\
\tilde{E}_{x} & 0 & \tilde{B}_{z} & -\tilde{B}_{y} \\
\tilde{E}_{y} & -\tilde{B}_{z} & 0 & \tilde{B}_{x} \\
\tilde{E}_{z} & \tilde{B}_{y} & -\tilde{B}_{x} & 0 \\
\end{bmatrix}
~.\label{FlatCovFab}
\end{equation}
we see that the Kasner $(1,0,0)$ electromagnetic fields are exactly the same 
as in Minkowski space, if we relate the Kasner fields ({\it before} the 
separation into non-static variables via 
Eq's.~\ref{SepVars}-\ref{GenKas2ndOrdkxkykz}) to Minkowski ones, 
by taking combinations like:
\begin{eqnarray}
\tilde{E}_{y} \equiv 
\frac{(E_{y} \tilde{t} + B_{z} \tilde{x})}{(\tilde{t}^{2} - \tilde{x}^{2})^{1/2}} 
= (E_{y} \cosh{x} + B_{z} \sinh{x}) ~, \nonumber \\
\tilde{B}_{z} \equiv 
\frac{(B_{z} \tilde{t} + E_{y} \tilde{x})}{(\tilde{t}^{2} - \tilde{x}^{2})^{1/2}} 
= (B_{z} \cosh{x} + E_{y} \sinh{x}) ~, \label{CombiningEMfields}
\end{eqnarray}
and so on.

Finally, note that the Kasner $(1,0,0)$ metric is not equivalent 
to the {\it entire} flat spacetime; but, as is obvious from 
Eq.~\ref{MinkEmTensorMixed} (and from the definitions 
in Eq.~\ref{FlatKas100revtransform}), the region of Minkowski 
space bounded by $|\tilde{t}| \geq |\tilde{x}|$, $t \geq 0$ 
is enough to cover the entire Kasner allowed coordinate range 
of $t_{\textrm{Kas}} \geq 0$, 
$-\infty < \{ x_{\textrm{Kas}}, y_{\textrm{Kas}}, z_{\textrm{Kas}} \} < \infty$.

\begin{acknowledgments}
I am grateful to Austin Nguyen for introducing me to the Julia 
programming environment that has been used for the calculations 
in this research; and to Elizabeth Wu for collaboration on 
preliminary numerical work regarding wave propagation 
restricted to each of the principal Kasner expansion axes.
%See Supplemental Material \cite{BochnerSupplementalCodes} 
%for copies of the codes used for these calculations. 
\end{acknowledgments}

\bibliography{Kasner_Paper--arXiv--Bochner}

\end{document}